\documentclass[a4paper,prd,superscriptaddress,11pt,noshowkeys,notitlepage,nofootinbib]{revtex4}
\usepackage[utf8]{inputenc}
\usepackage{ae,aecompl}
\usepackage{amsmath,amssymb,mathrsfs}
\usepackage{graphicx}
\usepackage{ccnishi}
\usepackage{dsfont}
\usepackage{slashed}
\usepackage{units}
\usepackage{nicefrac}

\usepackage[usenames,dvipsnames]{xcolor} 
\usepackage{hyperref}
\hypersetup{
    colorlinks=true,        
    linkcolor=purple,       
    citecolor=JungleGreen,  
    filecolor=magenta,      
}

\providecommand{\cM}{\mathscr{M}}

\providecommand{\cY}{\mathscr{Y}}
\providecommand{\cO}{\mathcal{O}}

\providecommand{\La}[1]{\Lambda_{#1}}

\providecommand{\tY}{\tilde{Y}}
\providecommand{\tw}{\tilde{w}}
\providecommand{\hw}{\hat{w}}
\providecommand{\btheta}{\bar{\theta}}

\providecommand{\ckmsm}{V_{\rm ckm}^{\rm sm}}

\providecommand{\hY}{\hat{Y}}
\providecommand{\ty}{\tilde{y}}
\providecommand{\tb}{\tilde{b}}

\providecommand{\lag}{\mathscr{L}}

\providecommand{\eps}{\epsilon}

\providecommand{\rank}{\operatorname{rank}}

\makeatletter
\newcommand*\bigcdot{\mathpalette\bigcdot@{.5}}
\newcommand*\bigcdot@[2]{\mathbin{\vcenter{\hbox{\scalebox{#2}{$\m@th#1\bullet$}}}}}
\makeatother

 \providecommand{\adriano}[1]{{\leavevmode\color{purple}#1}}

\begin{document}
\title{
Explicit parametrization of more than one vector-like quark of Nelson-Barr type
}
\author{G.~H.~S.~Alves}
\thanks{E-mail: alves.gustavo@aluno.ufabc.edu.br}
\affiliation{Centro de Ci\^{e}ncias Naturais e Humanas,
Universidade Federal do ABC, 09.210-170,
Santo Andr\'{e}-SP, Brasil}
\author{A.~L.~Cherchiglia}
\thanks{E-mail: alche@unicamp.br}
\affiliation{Instituto de Física Gleb Wataghin, Universidade Estadual de Campinas,
Rua Sérgio Buarque de Holanda, 777, Campinas, SP, Brasil}
\affiliation{Departamento de F\'isica Te\'orica y del Cosmos, Universidad de Granada, Campus de Fuentenueva, E–18071 Granada, Spain}
\author{C.~C.~Nishi}
\thanks{E-mail: celso.nishi@ufabc.edu.br}
\affiliation{Centro de Matem\'{a}tica, Computa\c{c}\~{a}o e Cogni\c{c}\~{a}o,
Universidade Federal do ABC, 09.210-170,
Santo Andr\'{e}-SP, Brasil}

\begin{abstract}
Nelson-Barr models solve the strong CP problem based on spontaneous CP violation and generically requires vector-like quarks (VLQs) mixing with standard quarks to transmit the CP violation.
We devise an explicit parametrization for the case of two VLQs of either down-type or up-type and quantitatively study several aspects including the hierarchy of the VLQ Yukawas and their irreducible contribution to $\btheta$.
In particular, with the use of the parametrization, we show that a big portion of the parameter space for two up-type VLQs at the TeV scale is still allowed by the constraint on $\btheta$, although this case had been previously shown to be very restricted based on estimates.
\end{abstract}
\maketitle
\section{Introduction}
\label{sec:intro}

The only source of CP violation in the SM measured so far is a single phase $\delta_{\rm CKM}$\,\cite{ParticleDataGroup:2022pth} residing in the Yukawa couplings of quarks and the Higgs, hence it manifests only in flavor violating phenomena.
The measurement of a similar phase in the leptonic sector is one of the major goals of planned neutrino oscillation experiments.
In contrast, a flavor conserving source of CP violation exists in principle in the SM in the form of the $\btheta$ term of QCD. However, the nonobservation of the eletric dipole moment of the neutron constrains this parameter to be tiny: $\btheta\lesssim 10^{-10}$\,\cite{EDMreview,nEDM:exp}. Why this is so constitutes the strong CP problem. 

The only known avenue to solve this problem without introducing an axion is to assume that CP\,\cite{strongCP:CP,nelson,barr} (or P\,\cite{strongCP:P}) is a fundamental symmetry of nature that is spontaneously broken.
The challenge is to arrange this breaking to generate an order one $\delta_{\rm CKM}$ but to suppress $\btheta$ sufficiently.
The best known examples with CP are based on the Nelson-Barr\,\cite{nelson,barr} idea which guarantees vanishing $\btheta$ at tree-level.
Then higher order corrections must be sufficiently suppressed. 
Corrections may already show up at one-loop\,\cite{BBP}, although it may be postponed to two-loops by using a nonconventional CP symmetry\,\cite{NB:CP4}.
Recent proposals based on spontaneously broken CP can be found in Refs.\,\cite{scpv:others:recent,hiller:01,meade:SFV,vecchi.2,perez.wise}.

In the Nelson-Barr setting, the spontaneous breaking of CP is transmitted to the SM through the mixing of SM quarks with heavy vector-like quarks (VLQs).
These were denoted as VLQs of Nelson-Barr type (NB-VLQs) in Refs.\,\cite{nb-vlq,nb-vlq:fit}.
It is possible to keep the typical corrections to $\btheta$ that appear at one- or two-loops under control if we suppress the couplings of the CP breaking scalars to these NB-VLQs or to the SM Higgs.
There are, however, corrections at 3-loops that cannot be arbitrarily suppressed because they depend only on the Yukawa couplings of NB-VLQs to the SM\,\cite{vecchi.1}, the same ones carrying the CP violation. 

Quite similarly to the axion solution, the Nelson-Barr solution also suffers from a quality problem\,\cite{dine,Asadi.Reece,choi.kaplan}
which requires that the CP breaking scale $\Lambda_{\rm CP}$ cannot be arbitrarily high. 
On the other hand, the stability of the domain walls from spontaneous breaking of the exact CP symmetry\,\cite{reece:stable}, requires that inflation takes place before CP breaking and this constraint leads to a number of cosmological implications\,\cite{Asadi.Reece}.
Additional gauge symmetries\,\cite{Asadi.Reece,perez.wise}, supersymmetry \,\cite{dine} or strong dynamics\,\cite{vecchi.2} can improve the quality and allow higher values for $\Lambda_{\rm CP}$. 

In contrast to the CP breaking sector, the NB-VLQs that mediate the CP breaking may lie at much lower energies, constrained to be above the TeV scale from collider searches.
To comply with the Barr criteria \cite{barr}, the NB-VLQs can only be electroweak singlets or doublets, in the same representations of SM quarks.
The case of doublets was argued to lead to too large corrections to $\btheta$\,\cite{vecchi.17}.
We have analyzed in Ref.\,\cite{nb-vlq} the case of one singlet by devising an explicit parametrization for the BSM parameters.
We have found that these VLQs typically couple to the SM quarks and Higgs following the hierarchy of the CKM last row or column, a feature that alleviates the strongest flavor constraints that apply to the first two quark families\,\cite{nb-vlq,nb-vlq:fit}.
In Ref.\,\cite{vecchi.1}, irreducible 3-loop contributions to $\btheta$ were analyzed through the construction of CP odd invariants and were estimated in terms of SM Yukawas and mixing.
It was shown that the case of two or more singlet up-type NB-VLQs was severely constrained by these contributions to $\btheta$.

Here we analyze in more detail the case of two or more singlet NB-VLQs of either down-type or up-type.
In particular, for two VLQs, we devise an explicit parametrization that allows us to explore the large parameter space.
In particular, the case of two NB-VLQs covers the case of vanishing $\btheta$ at one-loop protected by nonconventional CP\,\cite{NB:CP4}.
By being quantitative, we show that the VLQ Yukawa couplings typically follow the same hierarchies of the single VLQ case but with a possible variation that can be mapped out.
In particular, we show that the invariants estimating the 3-loop contributions to $\btheta$ for two NB-VLQs of up-type still allows for TeV scale VLQs.

The outline of this article is as follows.
In Sec.\,\ref{sec:model}, we review the model of singlet NB-VLQs. 
Section~\ref{sec:n=1} reviews the explicit parametrization for one NB-VLQ and analyzes several aspects of the model, including the case of special points where the VLQ Yukawa coupling may vanish for some flavors. 
In Sec.\,\ref{sec:n=2}, we describe the explicit parametrization for the case of two NB-VLQs and quantitatively study several aspects of the model, including the hierarchy of Yukawa couplings and heavy mass matrices.
The implications to the invariants of 3-loop contribution to $\btheta$ is also shown.
We summarize in Sec.\,\ref{sec:summary} and the appendices contain auxiliary results.

\section{Review of the model}
\label{sec:model}

We define down-type singlet VLQs of Nelson-Barr type (NB-VLQs) $B_{aL},B_{bR}$, $a,b=1,\dots,n_B$, through the lagrangian\,\cite{nb-vlq}
\eqali{
\label{yuk:NB}
-\lag&=\bar{q}_{iL}\cY^d_{ij} Hd_{jR}+\bar{q}_{iL}\cY^u_{ij} \tilde{H}u_{jR}
\cr
&\quad +
\bar{B}_{aL}\cM^{Bd}_{aj} d_{jR}+\bar{B}_{aL}\cM^B_{ab} B_{bR}+h.c.,
}
where they couple to the SM quark doublets $q_{iL}$ and singlets $d_{jR},u_{jR}$, $i,j=1,2,3$.
The definition requires that $\cY^u,\cY^d$ are \textit{real} $3\times 3$ matrices, $\cM_B$ is a real mass matrix and only $\cM^{Bd}$ is a complex matrix. 
This structure follows from CP conservation and a $\ZZ_2$ symmetry\,\footnote{a larger $\ZZ_n$ or $U(1)$ are also possible\,\cite{dine}.}
under which only $B_{L,R}$ are odd, and only $\cM^{Bd}$ breaks CP and $\ZZ_2$ softly (spontaneously) realizing the Nelson-Barr mechanism that guarantees $\bar{\theta}=0$ at tree-level\,\cite{nelson,barr}.
When not specified, we use the basis where $\cY^u=\hat{\cY}^u$ is diagonal, with a hat denoting diagonal matrices.

In contrast, \textit{generic} down-type VLQs are customarily described by the Lagrangian 
\eqali{
\label{yuk:VLQ}
-\lag&=\bar{q}_{iL}Y^d_{ij} Hd_{jR}+\bar{q}_{iL}Y^u_{ij} \tilde{H}u_{jR}
\cr &~~+\ 
\bar{q}_{iL} Y^B_{ia}H\,B_{bR} 
+ \bar{B}_{aL}M^B_{ab} B_{bR}+h.c.,
}
where $M_B$ is expected to be much larger than the electroweak scale.
We usually assume the basis where $Y^u=\hY^u$ is diagonal.
We have shown in Ref.\,\cite{nb-vlq} that \emph{one more} free parameter is needed compared to the case of one NB-VLQ. Hence, the NB case is just a subcase and when the lagrangian \eqref{yuk:NB} is rewritten in the form \eqref{yuk:VLQ}, the various parameters cannot be independent and correlations necessarily appear\,\cite{nb-vlq}.
In special, for one VLQ, \emph{only one} CP violating parameter controls all CP violation in the NB case while the generic case depends on three CP violating parameters\,\cite{lavoura.branco,branco:book}.

Similarly, singlet up-type NB-VLQs $T_{aL}, T_{bR}$, $a,b=1,\dots,n_T$, are defined by the lagrangian\,\cite{nb-vlq}
\eqali{
\label{yuk:NB:up}
-\lag&=\bar{q}_{iL}\cY^d_{ij} Hd_{jR}+\bar{q}_{iL}\cY^u_{ij} \tilde{H}u_{jR}
\cr
&\quad +
\bar{T}_{aL}\cM^{Tu}_{aj} u_{jR}+\bar{T}_{aL}\cM^T_{ab} T_{bR}+h.c.,
}
with only $\cM^{Tu}$ being complex,
while generic up-type VLQs can be described by
\eqali{
\label{yuk:VLQ:up}
-\lag&=\bar{q}_{iL}Y^d_{ij} Hd_{jR}+\bar{q}_{iL}Y^u_{ij} \tilde{H}u_{jR}
\cr &~~+\ 
\bar{q}_{iL} Y^T_{ir}\tilde{H}\,T_{aR} 
+ \bar{T}_{aL}M^T_{ab} T_{bR}+h.c.
}
Usually we assume the basis where $\cY^d=\hat{\cY}^d$ and $Y^d=\hY^d$. 
In most cases, to translate the result from down-type VLQs to up-type VLQs, we just need to relabel $d\leftrightarrow u$ and $B\to T$. 
So we will often omit the up-type case and write the explicit expressions only when necessary. When the distinction between down- and up-type is unimportant, we will also use $n$ instead of $n_B$ or $n_T$ for the number of VLQs.

The changing of basis from \eqref{yuk:NB} to \eqref{yuk:VLQ} can be performed analytically rotating only in the space $(d_R,B_R)$.
One simple choice leads to\,\footnote{
These relations were also given in Ref.\,\cite{vecchi.1}, except for the last one.
}
\subeqali[WR:YdYB]{
\label{Yd:NB:1/2}
Y^d&= \cY^d(\id_3-ww^\dag)^{+1/2}\,,
\\
\label{YB:cal-Yd}
Y^B&= \cY^dw\,,
\\
\label{WR:MB}
M^B&=\cM^B(\id_n-w^\dag w)^{-1/2}\,,
}
where 
\eq{
\label{def:w}
w=\tw(\id_n+\tw^\dag\tw)^{-1/2}\,,\quad
\tw^\dag = {\cM^B}^{-1}\cM^{Bd}\,.
}
See appendix \ref{ap:partial} for the details.
Changing basis from \eqref{yuk:NB:up} to \eqref{yuk:VLQ:up} is analogous after relabelling $d\to u, B\to T$ in \eqref{WR:YdYB}.
We can also write in implicit form,
\eq{
\label{wdagger}
w^\dag={M^B}^{-1}{\cM^{Bd}}\,.
}
For rotation in two dimensions, i.e., one SM family and one VLQ, we could write $w= \sin\theta$ while $\tw=\tan\theta$ for some angle $\theta$.

The relations \eqref{WR:YdYB} make more explicit the correlation between the SM Yukawa $Y^d$ and the VLQ Yukawa $Y^B$ for the case of NB-VLQs.
In the generic case, these Yukawas are independent but for the NB-VLQ case where \eqref{WR:YdYB} is valid, they  depend on common parameters with \emph{one less} parameter in total.

In order to eliminate the dependence on the unphysical rotation in $d_R$ space, we can rewrite the relation \eqref{Yd:NB:1/2} as 
\eq{
\label{Yd:NB}
Y^d{Y^d}^\dag =\cY^d\left(\id_3-ww^\dag\right){\cY^d}^\tp\,.
}
Therefore, in leading order, the latter is determined from SM input:
\eq{
\label{Yd:sm.input}
Y^d{Y^d}^\dag =V_{d_L}\diag(y^2_d,y^2_s,y^2_b)V_{d_L}^\dag\,,
}
where $V_{d_L}$ is the CKM matrix in the SM:
\eq{
\label{VdL:ckm}
V_{d_L}=\ckmsm\,,
}
in the basis where $Y^u$ is diagonal.
One can see that the only complex quantity in the righthandside of \eqref{Yd:NB} is $w$ (equivalently $\cM^{Bd}$) which should generate the CP violation in the CKM, typically requiring order one $w$\,\cite{nb-vlq,vecchi.1}.

For up-type VLQs, eq.\,\eqref{Yd:NB} is rewritten as
\eq{
\label{Yu:NB}
Y^u{Y^u}^\dag =\cY^u\left(\id_3-ww^\dag\right){\cY^u}^\tp\,.
}
while \eqref{Yd:sm.input} leads to
\eq{
\label{Yu:sm.input}
Y^u{Y^u}^\dag =V_{u_L}\diag(y^2_u,y^2_c,y^2_t)V_{u_L}^\dag\,.
}
The relation \eqref{VdL:ckm} between the diagonalization matrix and the CKM matrix is modified to
\eq{
\label{VuL:ckm}
V_{u_L}^\dag=\ckmsm\,,
}
in the basis where $Y^d$ is diagonal.

For NB-VLQs, given the structure of \eqref{Yd:NB} or \eqref{Yu:NB} which cannot be rephased, additional phases need to be taken into account in $V_{d_L}$ or $V_{u_L}$. So if we use a fixed parametrization for the CKM, we need to modify \eqref{VdL:ckm} or \eqref{VuL:ckm} to
\eq{
\label{phases.beta}
V_{d_L}=\mtrx{1&&\cr &e^{i\beta_1}& \cr && e^{i\beta_2}}\ckmsm\,,
\text{\quad or\quad}
V_{u_L}^\dag =\ckmsm \mtrx{1&&\cr &e^{-i\beta_1}& \cr && e^{-i\beta_2}}
\,.
}
We will often omit the phases $\beta_1,\beta_2$ when not relevant.
For generic VLQs, these phases can be transferred to the Yukawas $Y^B$ or $Y^T$ but for NB-VLQs the relation \eqref{YB:cal-Yd} forbids that.

Now let us use eq.\,\eqref{WR:YdYB} to relate the SM Yukawa with the VLQ Yukawa as
\eq{
Y^B=Y^d\tw\,.
}
Naively, it seems that $Y^B$ inherits the hierarchy of the SM Yukawa $Y^d$.
However, since the diagonalizing matrix from the right of $Y^d$ is unphysical, the relation is not so precise.
For one NB-VLQ of down type, its Yukawa couplings typically follow the hierarchy of the CKM\,\cite{nb-vlq}: 
\eq{
\label{typical.YB}
|Y^B_1|:|Y^B_2|:|Y^B_3|\sim |V_{ub}|:|V_{cb}|:|V_{tb}|
\sim 0.0036:0.04:1
\,.
}
For comparison, this is the exact hierarchy in the case where one VLQ couples exclusively with the third family.
For one up-type NB-VLQ, we analogously have 
\eq{
\label{typical.YT}
|Y^T_1|:|Y^T_2|:|Y^T_3|\sim |V_{td}|:|V_{ts}|:|V_{tb}|
\sim 0.0085:0.04:1
\,.
}
These properties roughly carry over to more than one NB-VLQs as we will see.
So this kind of hierarchy largely renders the model flavor safe as the most restrictive flavor constraints of flavor changing among the first and second families are naturally suppressed.
We should note however that the hierarchies \eqref{typical.YB} or \eqref{typical.YT} refer to typical values and they can be badly violated for special points\,\cite{nb-vll}.
We will also study how typical these properties are.

For future convenience, let us note that
\eq{
\label{YB:w-tilde}
{Y^d}^{-1}Y^B=\tilde{w}=\left({\cM^B}^{-1}\cM^{Bd}\right)^\dag
\,,
}
is the ratio between the CP violating contribution $\cM^{Bd}$ (mass) and the CP conserving bare mass $\cM^B$ of the NB-VLQs\,\cite{vecchi.1}.
In terms of the diagonalized Yukawa matrix
\eq{
Y^d=V_{d_L}\hY^d V_{d_R}^\dag\,,
}
we obtain
\eq{
\label{Yd-1YB}
{Y^d}^{-1}Y^B=V_{d_R}(\hY^d)^{-1}V_{d_L}^\dag Y^B=V_{d_R}(\hY^d)^{-1}\tY^B\,.
}
We have also incorporated the mixing matrix $V_{d_L}$ into $Y^{B}$ as
\eq{
\label{tildeYB}
\tY^B\equiv V_{d_L}^\dag Y^B\,,
}
where the Yukawa $Y^B$ is defined in the basis where $Y^u=\hY^u$. 
In the basis of trivial $V_{d_R}$, we can finally define
\eq{
\label{def:R}
R^d\equiv (\hY^d)^{-1}\tY^B\,.
}
This is the ratio between the Yukawas of the VLQ and the Yukawas of the SM quarks.
The analogous quantity for up-type NB-VLQs is
\eq{
\label{def:R:up}
R^u\equiv (\hY^u)^{-1}\tY^T\,,
}
where $\tY^T$ is analogous to \eqref{tildeYB} with $Y^T$ being the Yukawa in the basis where $Y^d=\hY^d$.
Note that the quantities \eqref{Yd-1YB} and \eqref{def:R} differ component-wise but their norms\,\footnote{For matrices, we use the Frobenius norm: $|A|=\sqrt{\Tr[AA^\dag]}$.} are the same:
\eq{
\label{R.norm}
|{Y^d}^{-1}Y^B|=|R^d|\,,\quad
|{Y^u}^{-1}Y^T|=|R^u|\,.
}
In the last equation, we showed the similar relation for up-type quantities.

At last, we can summarize the deviations that appear below the weak scale.
For both generic VLQs and NB-VLQs, we can use the same basis as \eqref{yuk:VLQ} and \eqref{yuk:VLQ:up} where a small mixing between $B_{aL}$ ($T_{aL}$) and $d_{iL}$ ($u_{iL}$) is induced by $Y^B$ ($Y^T$).
In leading order, the mixing matrix $V$ that appears in the couplings to $W$ is
\eqali{
\label{V:Theta}
\text{down-type VLQs:\quad} V&\approx \ckmsm
\left(\begin{array}{c|c}
\id_3-\ums{2}\Theta_d\Theta_d^\dag & \Theta_d
\end{array}\right)\,,
\cr
\text{up-type VLQs:\quad} V&\approx
\mtrx{
\id_3-\ums{2}\Theta_u\Theta_u^\dag 
\cr \Theta_u^\dag }
\ckmsm\,,
}
which are respectively of size $3\times (3+n)$ and $(3+n)\times 3$.
The coupling to $Z$ depends on $X^d=V^\dag V$ or $X^u=VV^\dag$.
At this order, all deviations depend on the mixing between SM quarks and VLQs, quantified by the matrices
\eqali{
\label{def:Theta}
\Theta_d &\equiv\frac{v}{\sqrt{2}}\tY^B {M^B}^{-1}\,,
\cr
\Theta_u &\equiv\frac{v}{\sqrt{2}}\tY^T {M^T}^{-1}\,.
}
If the VLQ mass matrix $M^B$ (or $M^T$) is not diagonal, further rotation on the spaces $B_{aL},B_{aR}$ ($T_{aL},T_{aR}$) might be necessary.

\section{One VLQ}
\label{sec:n=1}

Here we consider the case of one VLQ, which can be of down or up-type. 
For comparison with \cite{vecchi.1}, we will make use of the quantity $R^d$ in \eqref{def:R} or $R^u$ in \eqref{def:R:up}
which have size $R^d\sim R^u\sim 3\times 1$ for one VLQ. We recall that their norm is the ratio between the VLQ bare mass and the row matrix that control the mixing among VLQs and standard quarks; cf \eqref{YB:w-tilde}.

\subsection{Explicit parametrization}
\label{sec:param:n=1}

Here we briefly sketch the explicit parametrization found in Ref.\,\cite{nb-vlq}.
This parametrization manages to incorporate the 10  parameters of the SM flavor sector (6 quark masses and 4 parametrs in $\ckmsm$),
leaving 5 BSM parameters free.
The total number of parameters is 15.
For example, the Yukawa Lagrangian \eqref{yuk:NB} for one down-type NB-VLQ depends on 3 parameters in the up sector and 12 in the down sector. 
The up quark Yukawa couplings are determined and we only need to treat the 12 parameters of the down sector.

Focusing on one NB-VLQ of down-type, we basically need to invert \eqref{Yd:NB}
and find $\cY^d$ and $w$ in terms of $Y^d{Y^d}^\dag$. The latter is used as input, with 7 parameters fixed from the SM
whereas the phases $\beta_1,\beta_2$ in \eqref{phases.beta} are free.
For $w$ we choose the basis where
\eq{
\label{param:w}
w=\mtrx{0\cr ib\cr a}\,,
}
where $b<a$, $a^2+b^2<1$, with free $b\in [0,1/\sqrt{2}]$ and $a$ fixed by other parameters\,\cite{nb-vlq}.
The inversion is performed by
\eq{
\label{cal.Yd}
\cY^d=\Big[\re(Y^d{Y^d}^\dag)\Big]^{1/2}\cO \diag(1,\sqrt{1-b^2},\sqrt{1-a^2})\,,
}
where $\cO=(e_1|e_2|e_3)$ is a real orthogonal matrix determined from $Y^d {Y^d}^\dag$, with an additional free angle $\gamma$; the vector $e_1$ is the eigenvector of zero eigenvalue of $\La1^{-1/2}\im(Y^d {Y^d}^\dag)\La1^{-1/2}$ for $\La1=\re(Y^d{Y^d}^\dag)$.
We are left with the following 5 free parameters: 
\eqali{
\{M_B, b, \gamma, \beta_1, \beta_2\},
}
where $M_B$ is the VLQ mass.

\subsection{Perturbativity and CPV transmission}

Here we discuss the theoretical contraints on the VLQ Yukawa couplings\,\cite{nb-vlq,vecchi.1}.
On the one hand, they cannot be arbitrarily large if we require perturbativity.
On the other hand, they cannot be arbitrarily small as the transmission of the CP violation to the CKM matrix depends on them.

To be explicit, we require for perturbativity, 
\eq{
\label{perturbativity:n=1}
|\tY^B_i|<4\pi\,,\quad |\tY^T_i|<4\pi\,,
}
for each component. For two or more VLQs, these relations are extended to all components:
\eq{
\label{perturbativity:n=2}
|\tY^B_{ia}|<4\pi\,,\quad |\tY^T_{ia}|<4\pi\,,
}
where $a=1,\dots,n$ runs over the VLQs.

\begin{figure}[h]
\includegraphics[scale=0.8]{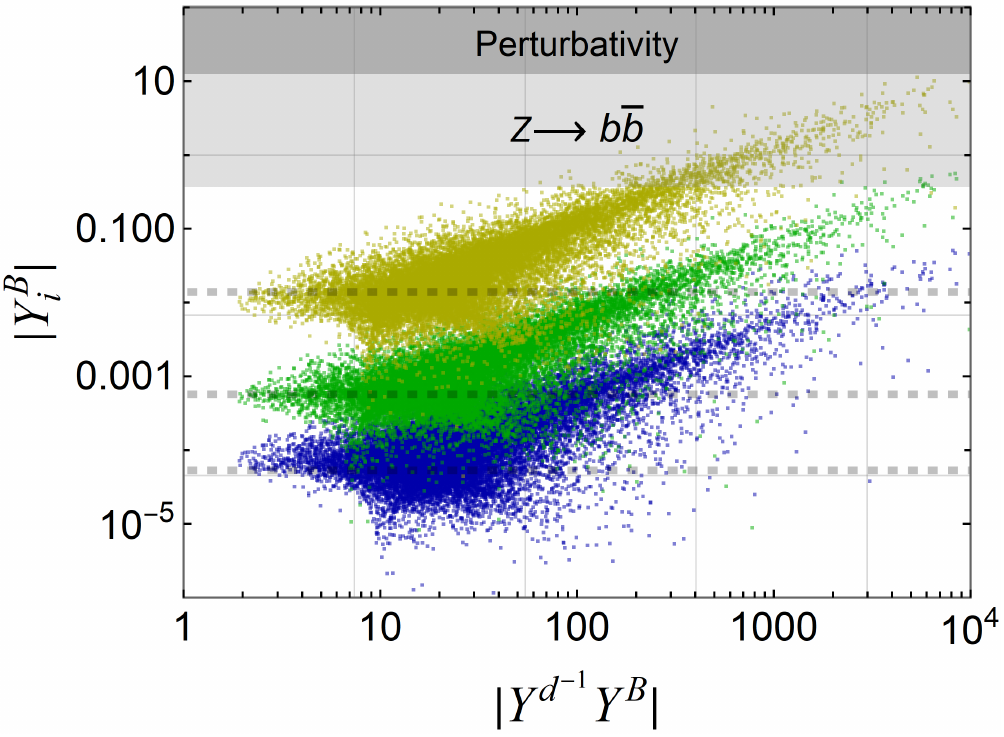}
\includegraphics[scale=0.785]{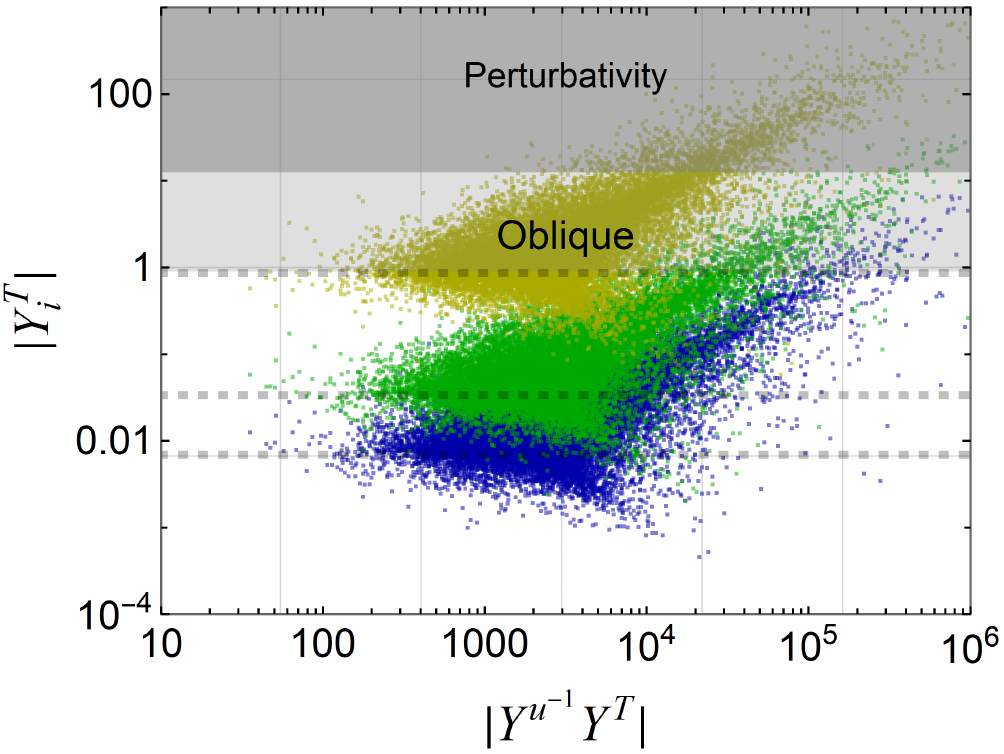}
\caption{\label{fig:YQi:n=1}
Components ($i=3,2,1$, in yellow, green and blue respectively) of the VLQ Yukawas for $n=1$ as a function of $|R^d|=|{Y^d}^{-1}Y^B|$ or $|R^u|=|{Y^u}^{-1}Y^T|$.
The dashed lines show $y_b|V_{ib}|$, $i=t,c,u$ (left) and $y_t|V_{ti}|\sim |V_{ti}|$, $i=b,s,d$ (right). The dark shaded areas denote the perturbativity constraint \eqref{perturbativity:n=1}. The light shaded areas are valid for $M_B=1.2\,\unit{TeV}$ and $M_T=1.3\,\unit{TeV}$, and they go up inversely proportional with the mass; see Sec.\,\ref{sec:flavor:n=1}.
}
\end{figure}

Let us see how the perturbativity constraint affects the quantity $R^d$ in \eqref{def:R} or $R^u$ in \eqref{def:R:up}, noting that their norms are the same as the norms of ${Y^d}^{-1}Y^B$ or ${Y^u}^{-1}Y^B$, respectively; cf.\,\eqref{R.norm}.
To analyze that, we show in Fig.\,\ref{fig:YQi:n=1} (left) the components $|Y^B_i|$ as a function of $|R^d|=|{Y^d}^{-1}Y^B|$. The up-type case is shown on the right figure.
In dashed lines we show $y_b|V_{ib}|$, $i=t,c,u$, and we clearly see that the Yukawa couplings $Y^B_i$ typically obey the hierarchy \eqref{typical.YB} of the third column of the CKM matrix.
Analogously, on the right, we see the hierarchy  
\eqref{typical.YT} of the third row of the CKM matrix. The dark shaded area denotes the perturbativity constraint \eqref{perturbativity:n=1} applied to $Y^B_i$ instead of $\tY^B$ and analogously for $Y^T_i$.
We see that such a constraint is relevant only for the up-type VLQ.

The perturbatitivy constraint \eqref{perturbativity:n=1} implies for the the third component
\eq{
\label{Rd3}
|R^d_3|=\left|\frac{1}{y_b}\tY^B_3\right|\lesssim \frac{4\pi}{y_b}\sim 900\,,
}
where we use the running Yukawa couplings of the SM at the TeV scale\,\cite{antusch}.
Since $\tY^B_i$ is just $Y^B_i$ rotated by the CKM, 
the typical hierarchy of $\tY^B_i$ will be roughly the same as of \eqref{typical.YB}.
Considering that the hierarchy of the down quark Yukawas are stronger than $|V_{ib}|$, the constraint on \eqref{Rd3} will be the strongest.
To be more precise, we will see in Sec.\,\ref{sec:Ri:n=1} that the typical hierarhcy for the components of $R^d$ will be $R^d\sim (10,2,1)$.
Then we can convert the constraint on the norm
\eq{
\label{Rd:n=1:pert}
|R^d|\sim \sqrt{105}|R^d_3|\lesssim 9200\,.
}
This is roughly the value in  Fig.\,\ref{fig:YQi:n=1} attained by $|{Y^d}^{-1}Y^B|$ when $|Y^B_3|$ reaches the perturbativity limit (dark shaded region).

Similarly, for an up-type VLQ, 
\eq{
|R^u_3|=\left|\frac{1}{y_t}\tY^T_3\right|<\frac{4\pi}{y_t}\sim 14\,.
}
The typical hierarchy $R^u\sim (1000,10,1)$, studied in the next section, leads to
\eq{
\label{Ru:n=1:pert}
|R^u|\lesssim 1.4\times 10^4.
}
This is also roughly the value in  Fig.\,\ref{fig:YQi:n=1} attained by $|{Y^u}^{-1}Y^T|$ when $|Y^T_3|$ reaches the perturbativity limit (dark shaded region).

Now, let us turn to the lower bound on $|R^d|$ (or $|R^u|$).
It comes from the requirement that the imaginary part of $\tw={Y^d}^{-1}Y^B$ must attain a minimum value in order to allow the correct generation of complex CKM.
This can be understood by rewriting \eqref{Yd:NB} as 
\eq{
{\cY^d}^{-1}Y^d{Y^d}^\dag{\cY^d}^{\tp-1}
=\id_3-ww^\dag=(\id_3+\tw\tw^\dag)^{-1}\,,
}
that is, 
\eq{
{\cY^{d}}^\tp(Y^d{Y^d}^\dag)^{-1}\cY^d-\id_3
=\tw\tw^\dag\,.
}
One can never eliminate completely the lefthandside by multiplying the \emph{real} matrix $\cY^d$ to the \emph{complex} matrix $(Y^d{Y^d}^\dag)^{-1}$. So there must be a minimum value for the norm of $\tw$ on the righthandside.
In Ref.\,\cite{nb-vlq}, we have used a similar reasoning to extract the minimum amount of FCNC. 
In Ref.\,\cite{vecchi.1}, this constraint was estimated to extract the minimum value for $|\tw|=|R^d|$.

Here, instead, we use our parametrization to directly extract the lower bound from Fig.\,\ref{fig:YQi:n=1}.
We also update the estimates for the perturbativity bounds in \eqref{Rd:n=1:pert} or \eqref{Ru:n=1:pert} from the figure, noting that some variation beyond the typical behavior can be clearly seen.
Collecting these bounds from the figure, we obtain
\eqali{
\label{range:R}
2&\lesssim |R^d|\lesssim 10^4\,,
\cr
30&\lesssim |R^u|\lesssim 3\times 10^5\,.
}
These values differ from Ref.\,\cite{vecchi.1} even if we translate them from component to norm and also correct for the different hierarchy we find for $|R_i|$ in \eqref{R:hierarchy}.
This difference stems in part from the use of our general parametrization which captures in more detail the possible variation of the parameters.

\subsection{Hierarchy of $R_i$}
\label{sec:Ri:n=1}

Here we study the hierarchy of the quantity $R_i$ defined in \eqref{def:R} or in \eqref{def:R:up}.
They depend on the known Yukawa couplings $\hY^d$ or $\hY^u$, and on the unknown Yukawa couplings $Y^B_i$ or $Y^T_i$.

\begin{figure}[h]
\includegraphics[scale=0.8]{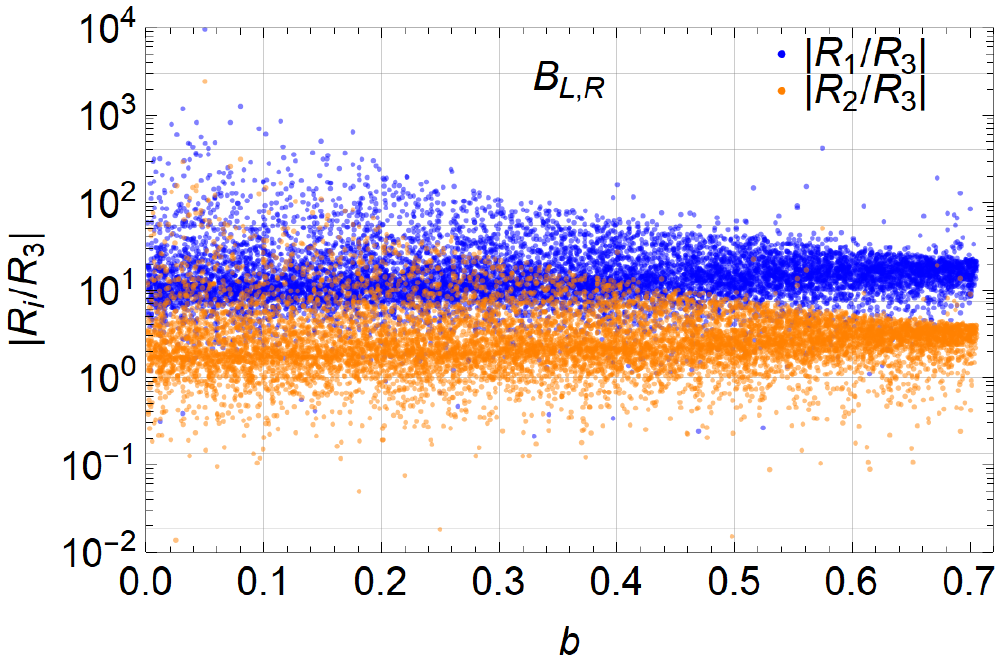}
\includegraphics[scale=0.8]{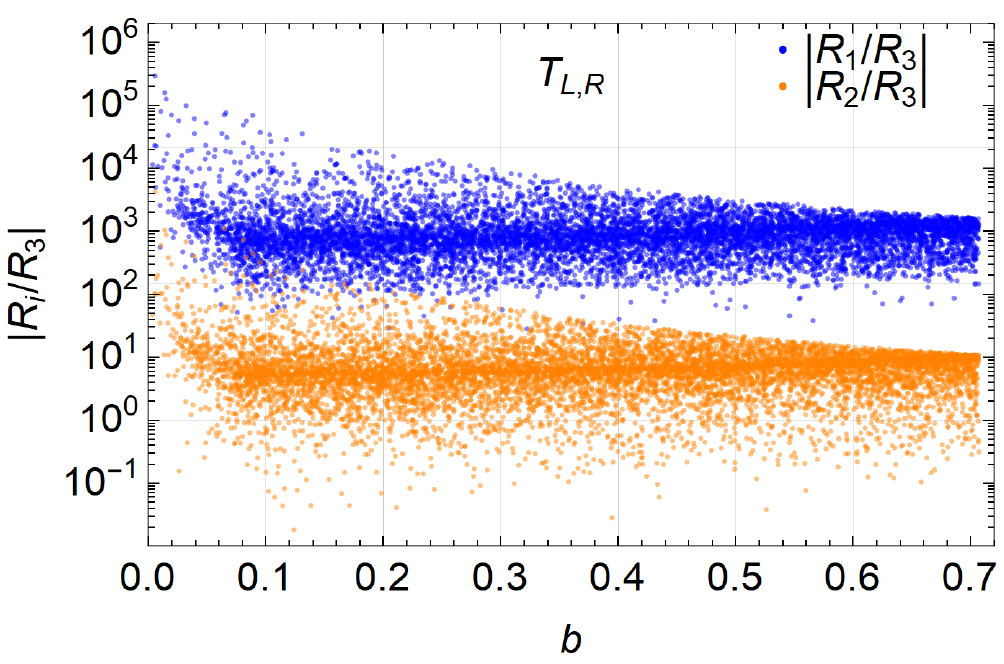}
\caption{\label{fig:ratios:n=1}
Ratios of $R_i/R_3$ in \eqref{def:R} quantifying the strength of the Yukawa coupling to the NB-VLQ compared to the SM Yukawas.
}
\end{figure}

The VLQ Yukawa couplings with the SM quarks, $Y^B$ or $Y^T$, was shown in the previous section to typically follow the hierarchy of the CKM matrix in \eqref{typical.YB} or \eqref{typical.YT}.
However, the deviation from these typical values may be larger than an order of magnitude, so that simple estimates of the hierarchy for $R_i$ may not be reliable.
So we use the explicit parametrization described in Sec.\,\ref{sec:param:n=1} to study the hierarchy of $R_i$.
In Fig.\,\ref{fig:ratios:n=1} we show randomly generate points of the ratio $R_{i}/R_{3}$ against the parameter $b$.
Typically, with a large variation, we see a mild hierarchy for a down-type NB-VLQ and a larger hierarchy for an up-type NB-VLQ:
\eqali{
\label{R:hierarchy}
|R^d_1|:|R^d_2|:|R^d_3| &\approx 10:2:1\,,\cr
|R^u_1|:|R^u_2|:|R^u_3| &\approx 10^3:10:1\,.
}
Notice that the definition \eqref{def:R} or \eqref{def:R:up}, combined with the typical hierarchy
\eqref{typical.YB} or \eqref{typical.YT}, leads to similar values for the up-type case but it underestimates $|R^d_1|$. 

\begin{figure}[h]
\includegraphics[scale=0.8]{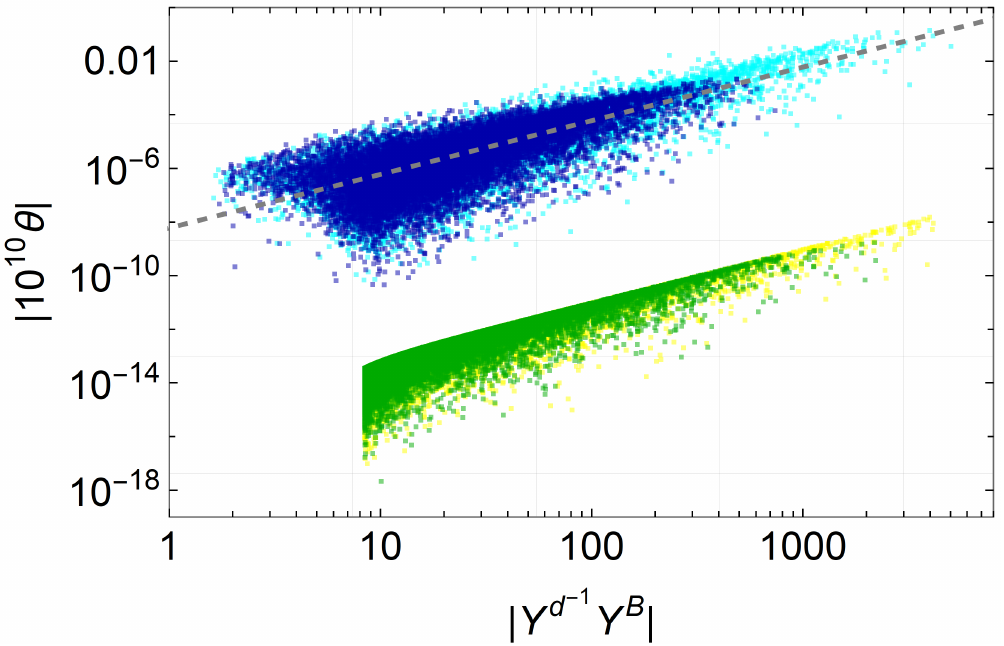}
\includegraphics[scale=0.8]{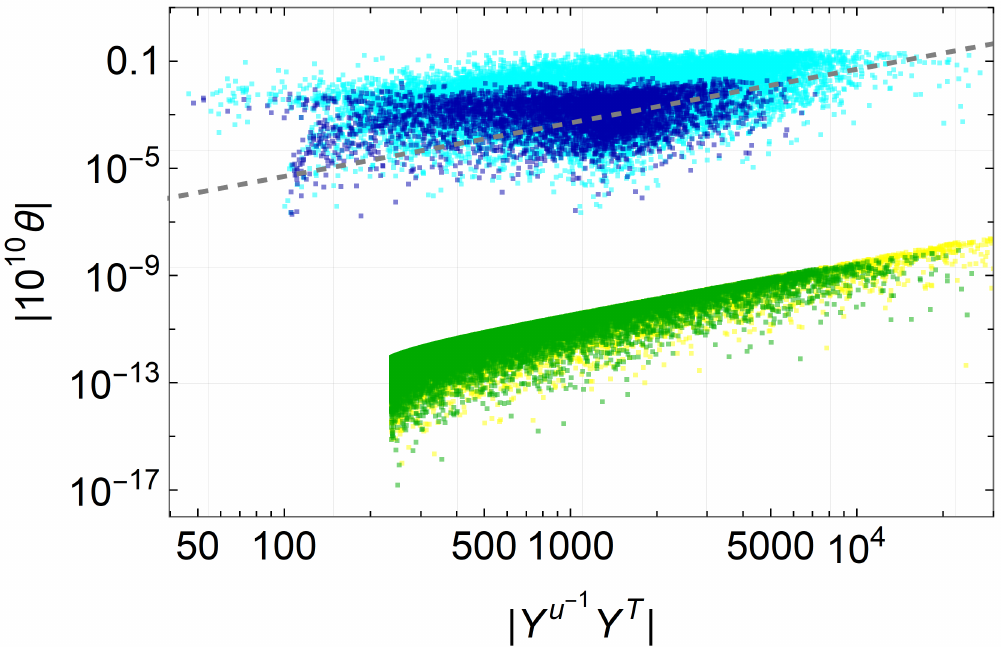}
\caption{\label{fig:theta:n=1}
Dominant invariant contributing to $\bar{\theta}$ as a function of $|(Y^q)^{-1}Y^Q|=|R^q|$ for NB-VLQ of down ($q=d,Q=B$) and up ($q=u,Q=T$) types.
The dark blue is for $M_B=1.2\,\unit{TeV}$ ($M_T=1.3\,\unit{TeV}$) while the light blue is for $M_B=12\,\unit{TeV}$ ($M_T=13\,\unit{TeV}$). 
The green and yellow points refer to special points, cf.\,Sec.\,\ref{sec:special}, with the same light/heavy masses.
}
\end{figure}

\subsection{Invariants estimating $\bar{\theta}$}
\label{sec:invariants:n=1}

The CP violation of a model can be quantified through invariants that are defined by a combination of the CP violation sources of that model and are independent of the reparametrization of their fields. In the SM, the unique experimentally measured CP violation is described by the Jarlskog invariant \cite{Jarlskog}.

In Nelson-Barr models, $\btheta$ is arranged to vanish at tree level\,\cite{nelson,barr} but loop corrections may arise already at the 1-loop level\,\cite{BBP}.
These corrections, however, may be arbitrarily suppressed by suppressing one or both of the following couplings: the Yukawa coupling between SM quarks and VLQs or the scalar portal couplings between the CP violating scalar(s) and the SM Higgs.

However, we have shown in \cite{nb-vlq} that the Yukawa couplings between the SM quarks and the VLQs cannot be arbitrarily small as full decoupling of the VLQs would not provide the SM with the necessary CP violation.
Therefore, as shown in Ref.\,\cite{vecchi.1}, there are irreducible and non-decoupling contributions, first appearing at 3-loops, that depends solely on these Yukawa couplings.
These contributions were estimated through the construction of CP odd flavor invariants.
For a single down-type and a single up-type VLQ, the leading order invariants with fewer SM Yukawas insertions were shown to be\,\cite{vecchi.1} 
\eqali{
\label{n=1:invariants}
I^d_{1,0}&\equiv \left( {Y^B}^\dag\left[ Y^d{Y^{d}}^\dagger, Y^u{Y^{u}}^\dagger\right]Y^B\right),
\cr
I^u_{1,0}&\equiv \left( {Y^T}^\dag\left[ Y^d{Y^{d}}^\dagger, Y^u{Y^{u}}^\dagger\right]Y^T\right)
\,.
}
These invariants provide an estimate for $\btheta$ within an order of magnitude since order one pre-factors are expected in a calculation within a full model.

Let us review the estimates given in Ref.\,\cite{vecchi.1} for the 3-loop contribution to $\btheta$ coming from the invariants \eqref{n=1:invariants}. For a single down-type VLQ, the estimate is
\eqali{
\label{theta:d:n=1}
\bar{\theta}&\sim \left(\frac{1}{16 \pi^2}\right)^3 \text{Im}(I^d_{1,0}),
\cr
&\sim \frac{\lambda^2_C y^2_t y^3_b y_s}{\left(16 \pi^2\right)^3} \frac{|R^d|^2}{3}\,,
\cr
&\approx 6 \times 10^{-18} \frac{|R^d|^2}{3}\,,
}
where $\lambda_C\sim 0.23$,
while for a single up-type VLQ, the estimate is
\eqali{
\label{theta:u:n=1}
\bar{\theta} &\sim \left(\frac{1}{16 \pi^2}\right)^3 \text{Im}(I^u_{1,0}),
\cr
&\sim \frac{\lambda^2_Cy^2_b y^3_ty_c}{\left(16 \pi^2\right)^3}\text{Im}(R^u_2R^{u^*}_3),
\cr
&\sim 5 \times 10^{-15} \lambda_C^2\frac{|R^u|^2}{2}.
}
We are including the factor $1/3$ in \eqref{theta:d:n=1} because Ref.\,\cite{vecchi.1} considers that $R^d_i$ have all the components of the same order, i.e,
\eq{
\label{Rd:ratios:vecchi}
|R^d_1|:|R^d_2|:|R^d_3|\approx 1:1:1\,,
}
and considers the dependence with respect to one of these generic components while we are considering the dependence on the norm $|R^d|$.
Similarly, in \eqref{theta:u:n=1}, the factors $\lambda_C^2/2$ appears because Ref.\,\cite{vecchi.1} considers
\eq{
\label{Ru:ratios:vecchi}
|R^u_1|:|R^u_2|:|R^u_3|\approx 1:1:\lambda_C^2\,,
}
and we are adapting the dependence on the component $|R^u_2|$ to the norm $|R^u|$.

However, by using our parametrization of Sec.\,\ref{sec:param:n=1}, we can generate points to test the properties \eqref{Rd:ratios:vecchi}
and \eqref{Ru:ratios:vecchi}.
The result, which is shown in Fig.\,\ref{fig:ratios:n=1}, clearly demonstrates that the ratios $|R_i|/|R_3|$ are typically very different.
We can see that typically the ratios are closer to \eqref{R:hierarchy}.

If we adopt these values in the estimates of \eqref{theta:d:n=1} and \eqref{theta:u:n=1}, we obtain
\eqali{
\left(\frac{1}{16 \pi^2}\right)^3 \im(I^d_{1,0})
&\approx 6 \times 10^{-18} \frac{|R^d|^2}{10}\,,
\cr
\left(\frac{1}{16 \pi^2}\right)^3 \text{Im}(I^u_{1,0})
&\approx 5 \times 10^{-20}|R^u|^2.
}
We see that the estimate of $\im(I^d_{1,0})$ is corrected by an order one factor but, in contrast, $\im(I^u_{1,0})$ is corrected by a very suppressed factor $1/5300$.

In fact, one can see in Fig.\,\ref{fig:theta:n=1} that these corrected estimates are in excellent agreement with the randomly generated points. 
The figure shows $10^{10}\btheta$ as a function of $|R^d|$ or $|R^u|$ and these estimates can be seen in the dashed lines. The randomly generated points are given in the dark blue ($M\sim \unit{TeV}$) and light blue ($M\sim 10\,\unit{TeV}$).

\subsection{Flavor and electroweak constraints}
\label{sec:flavor:n=1}

In general, the presence of VLQs induce modifications of flavour observables. Particularly relevant are FCNCs generated by the $Z$ exchange, and the inclusion of VLQs in box diagrams due to their mixing with the SM quarks. Therefore, there are stringent bounds that models with VLQs must pass in general. In the case of VLQs of down-type, a global fit was performed rendering allowed regions for the products of Yukawas couplings connecting the VLQ and the SM quarks  \cite{nb-vlq:fit}. As a by-product, it was also possible to extract upper limits for the transition between the VLQ and up quarks induced by the $W$ boson as below
\eqali{
\label{flavor:B}
|V_{tB}|\leq0.054,
\cr
|V_{cB}|\leq0.0086,
\cr
|V_{uB}|\leq0.013.
}
We will use these bounds in our plots, noting that on the Yukawas, they vary linearly with the VLQ mass as
\eq{
V_{iB}\approx \frac{v}{\sqrt{2}M_B}Y^B_i\,,
}
at leading order; cf.\,\eqref{V:Theta}.

Regarding VLQs of up-type, similar bounds can be extracted. For the transition between the VLQ and the bottom induced by the $W$ boson we will use the constraints established in \cite{Saavedra 2013}, where only mixing with third family SM quarks was considered. For the product of Yukawa couplings, bounds in terms of the VLQ mass can be found in \cite{IshiwataLigetiWise2015} where a plethora of flavour observables were analised. Thus, in our plots we will use the following constraints  
\eqali{
\label{flavor:T}
|V_{Tb}|\leq0.12,
\cr
|Y_3^T Y_1^T|\leq \frac{M_T}{25\;\unit{TeV}},
\cr
|Y_3^T Y_2^T|\leq \frac{M_T}{6.4\;\unit{TeV}},
\cr
|\text{Re}(Y_1^{T^*} Y_2^T)|\leq \frac{M_T}{42\;\unit{TeV}},
\cr
|\text{Im}(Y_1^{T^*} Y_2^T)|\leq \frac{M_T}{670\;\unit{TeV}}\,.
}

Both constraints \eqref{flavor:B} and \eqref{flavor:T} are valid for VLQs at the TeV scale.\footnote{%
The first constraint in \eqref{flavor:T} comes from oblique parameters $S,T$\,\cite{Saavedra 2013} for $M_T=1.3\,\unit{TeV}$. For higher masses the constraint is tighter but we neglect this weak variation.
}
For VLQs with masses around $7\,\unit{TeV}$ or heavier, box contributions with Higgs exchange start to dominate as they lead to a different scaling in the mass as $Y^4/M$\,\cite{IshiwataLigetiWise2015} for both $B$ or $T$ VLQs.

\subsection{Special points}
\label{sec:special}

The NB-VLQs cannot decouple from the SM as they need to transmit the spontaneous CP violation to the other sectors of the model. However, it is possible to eliminate the coupling of NB-VLQ with a specific SM quark flavor for special choices of the phases $\{\beta_1,\beta_2\}$ in \eqref{phases.beta}, appearing in CKM rephasing convention.\,\footnote{Unlike in the SM, these phases are physical in the presence of VLQs.}
This property was proved in Ref.\,\cite{nb-vll} for the similar case of vector-like leptons transmitting CP violation to the leptonic sector of the SM.
For the case of one NB-VLQ, these special choices lead to one of the following patterns:
\eqali{
\label{special:pattern}
(\tilde{Y}^B_d,\tilde{Y}^B_s,\tilde{Y}^B_b)&=(0,\times,\times)\quad \text{or}\quad (\times,0,\times)\quad \text{or} \quad (\times,\times,0),
\cr
(\tilde{Y}^T_u,\tilde{Y}^T_c,\tilde{Y}^T_t)&=(0,\times,\times)\quad \text{or}\quad (\times,0,\times)\quad \text{or} \quad (\times,\times,0).
}
By choosing one more free parameter appropriately, we can further make two components of $\tY^B$ or $\tY^T$ vanish\,\cite{nb-vll}. We will not treat this subcase here.

For the case of one down-type (up-type) VLQ, the matrix $V_{d_L}$ ($V_{u_L}$) is the matrix that diagonalizes $Y^d{Y^d}^\dag$ ($Y^u{Y^u}^\dag$); see eq.\,\eqref{Yd:sm.input}.
In the basis where $Y^u$ is diagonal for the down-type VLQ (or $Y^d$ is diagonal for the up-type VLQ), these diagonalizing matrices are related to the CKM matrix of the SM as
\eq{
V_{d_L}=\ckmsm=(v_1|v_2|v_3),
}
or
\eq{
V_{u_L}^\dag=\ckmsm=(v_1|v_2|v_3)^\dag\,.
}
We see that the vectors $v_i$ are the eigenvectors of the matrix $Y^d{Y^d}^\dagger$ or $Y^u{Y^u}^\dagger$ in each case.

By rephasing rows (down-type) or columns (up-type) of $\ckmsm$ we can always make one of the columns or rows $v_i$ real.
This is equivalent to choosing appropriate phases $\beta_i$ in \eqref{phases.beta}.

For illustration, to show the third pattern in \eqref{special:pattern}, i.e., $(\tY^B)_3=v_3^\dag Y^B=0$, we need to show that $Y^B$ is orthogonal to $v_3$.
Let us choose $v_3$ real, corresponding to having the third row of the CKM matrix real. This implies that
\eq{
\label{ReImYdYd}
\text{Re}(Y^d{Y^{d}}^\dagger)v_3=y^2_3v_3\,, \quad  \text{Im}(Y^d{Y^{d}}^\dagger)v_3=0\,.
}
As a result, using the notation $\re(Y^d{Y^{d}}^\dagger)=\La1 $ and $\im(Y^d{Y^{d}}^\dagger)=\La2$, the real vector $v_3$ is also an eigenvector of $\La1^{-1/2}\La2\La1^{-1/2}$ with eigenvalue zero.
So the first column $e_1$ of the orthogonal matrix $\cO$ in \eqref{cal.Yd} is either $v_3$ or $-v_3$.
Then 
\eq{
v^\dagger_3\cY^d=\pm y_3(1,0,0)\,.
}
Since $Y^B=\cY^d w$ and $w$ in \eqref{param:w} is orthogonal to the vector above, we obtain the desired property.

Although we can choose to eliminate other components of $\tY^B$, quantitatively, this is the case where we can see the most significant difference because $\tY^B_3$ tends to be hierarchically larger for generic points.
In Fig.\,\ref{fig:theta:n=1} (left) we show these special points with $\tY^B_3=0$ in green ($M_B=1.2\,\unit{TeV}$) and yellow ($M_B=12\,\unit{TeV}$).
We clearly see the values of the estimate of $\btheta$ are highly suppressed compared to the generic case (blue and cyan).
On the right we also show the analogous case of $\tY^T_3=0$ for similar value of masses and the suppression is similar.
This strong suppression is sufficient to evade any experimental bound on $\btheta$.

\section{Two or more VLQs}
\label{sec:n=2}

We focus mainly on the case of $n=2$ VLQs of either down or up type but we treat the case of any $n$ when possible.

We will detail the parametrization for $n=2$ in the following subsections but we anticipate some results based on it.

\begin{figure}[h]
\includegraphics[scale=0.8]{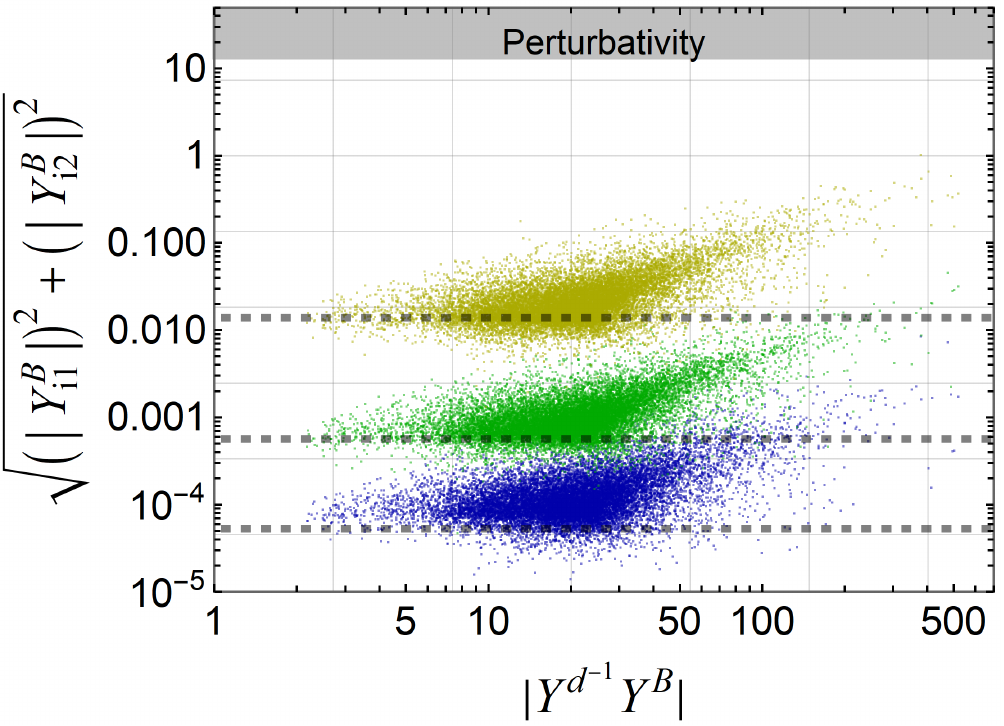}
\includegraphics[scale=0.78]{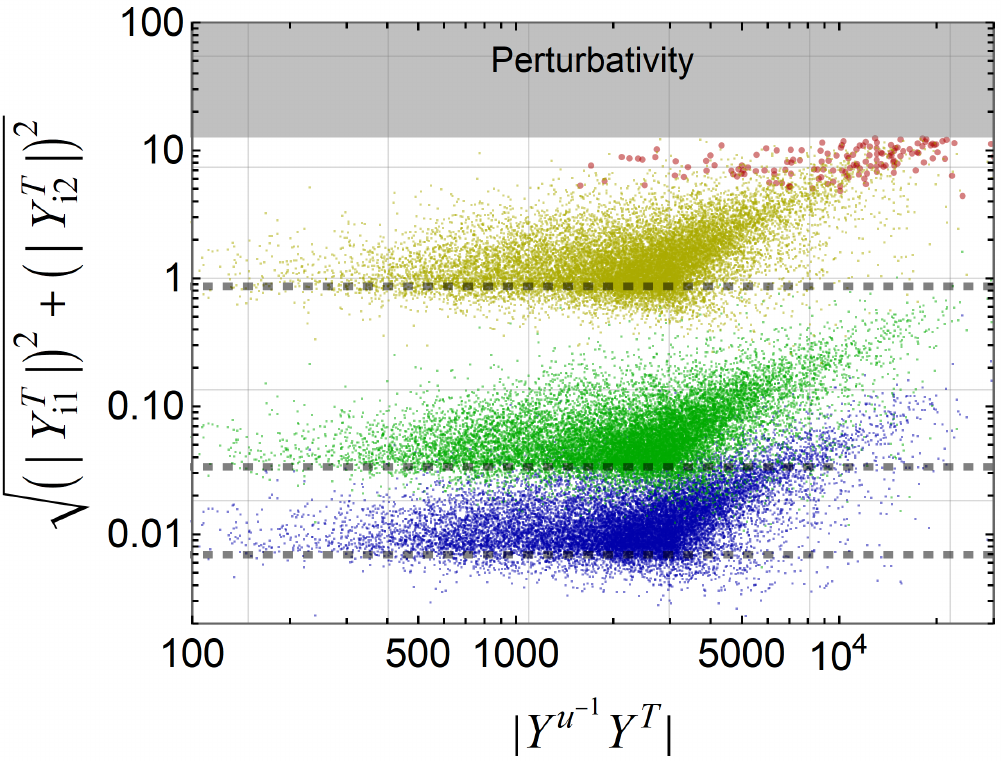}
\caption{\label{fig:YQ:n=2}
Left: norm $|Y^B_i|$, cf.\,\eqref{YBi:norm}, as a function of the norm $|{Y^d}^{-1}Y^B|=|R^d|$ for $i=3,2,1,$ respectively in yellow, green and blue. 
Right: similar plot for two up-type NB-VLQs. 
The dashed lines show $y_b|V_{ib}|$, $i=t,c,u$ (left) and $y_t|V_{ti}|\sim |V_{ti}|$, $i=b,s,d$ (right). The dark shaded areas denote the perturbativity constraint \eqref{perturbativity:n=2}. 
The red points (right) are filtered with electroweak $S,T$; see Sec.\,\ref{sec:flavor:n=2}.
}
\end{figure}

Firstly, Fig.\,\ref{fig:YQ:n=2} (left) shows the norm
\eq{
\label{YBi:norm}
|Y^B_i|\equiv \sqrt{|Y^B_{i1}|^2+|Y^B_{i2}|^2}\,,
}
as a function of $|{Y^d}^{-1}Y^B|=|R^d|=\sqrt{\sum_{ia}|R^d_{ia}|^2}$.
The figure in the right shows the same plot for two up-type VLQs.
For both cases, we clearly see a hierarchy of the components that couple to each $q_{iL}$ following the hierarchy of the CKM in \eqref{typical.YB} or \eqref{typical.YT}, which are shown in dashed lines. 
The typical Yukawa couplings are hierarchically larger for the couplings with the heavier SM quark.
The perturbativity constraint \eqref{perturbativity:n=2} is shown in the dark gray band. 
The red points are the ones filtered using the dominant flavor or electroweak constraint, as explained in Sec.\,\ref{sec:flavor:n=2}. 
The mass spectrum of the VLQs follow the $\kappa=1$ case of Fig.\,\ref{fig:MB:n=2:k=1,1-100}, with $\mu_0=1.2\,\unit{TeV}$ ($\mu_0=1.3\,\unit{TeV}$) for down-type (up-type) VLQs. These values correspond roughly to the lightest mass. The values for the Yukawa coupligns, however, do not change significantly if we change the spectrum ($\mu_0,\kappa$).

\begin{figure}[h]
\includegraphics[scale=0.81]{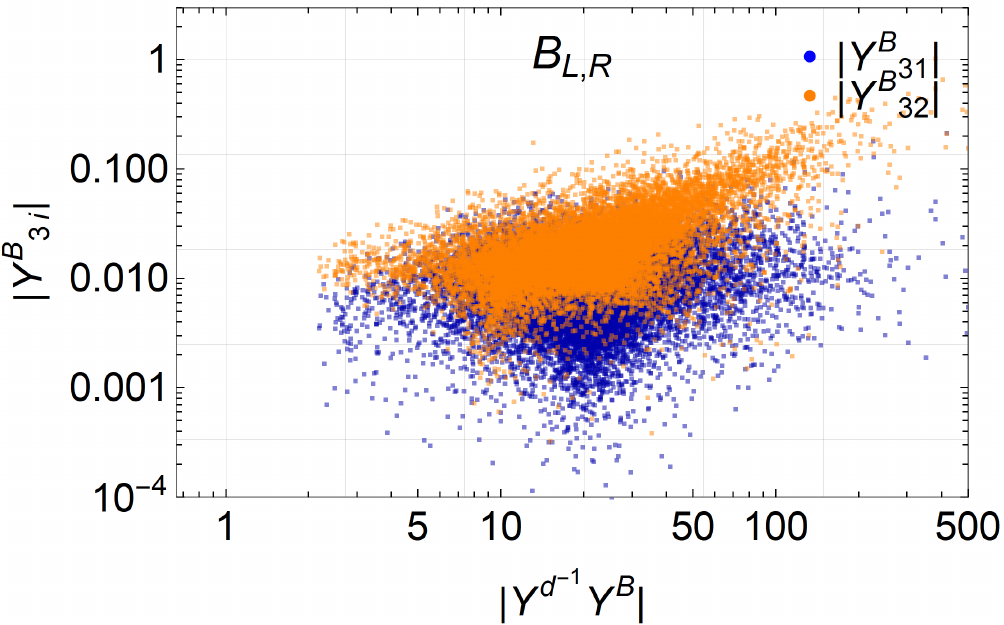}
\includegraphics[scale=0.76]{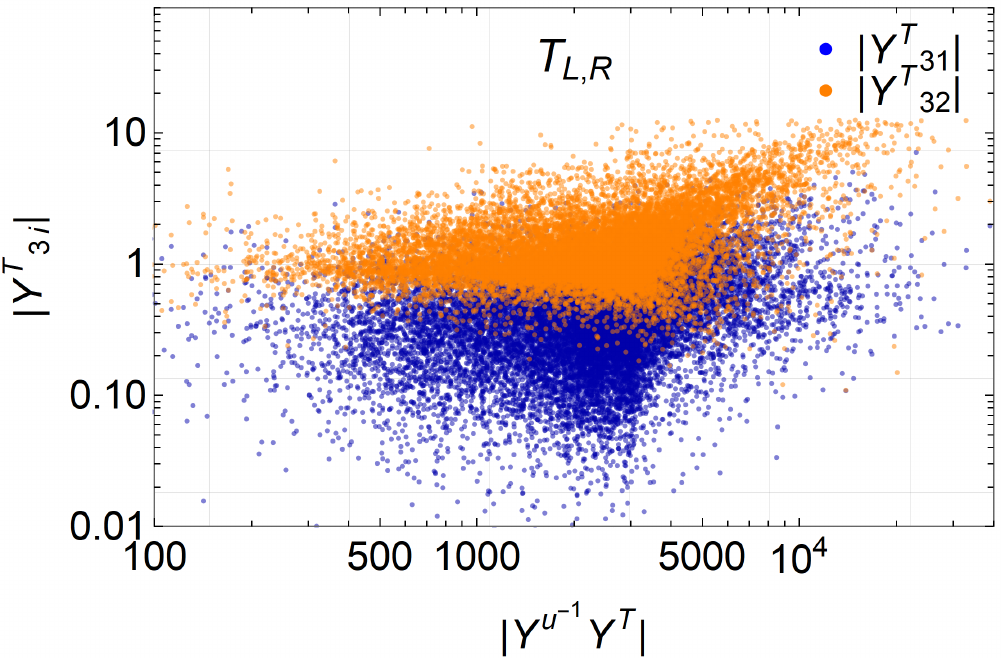}
\caption{\label{fig:Y31Y32:n=2}
Left: plot for $|Y^B_{31}|$ and $|Y^B_{32}|$ as a function of the norm $|{Y^d}^{-1}Y^B|=|R^d|$.
Right: similar plot for two up-type NB-VLQs. 
}
\end{figure}

To analyze the relative size of the Yukawa couplings $Y^B_{3a}$ to $B_a$, $a=1,2$, we show in Fig.\,\ref{fig:Y31Y32:n=2} (left) the values of $|Y^B_{31}|$ (blue) and $|Y^B_{32}|$ (orange) as a function of $|R^d|$.
Note that we conventionally order $B_1,B_2$ from lighter to heavier.
We see that the Yukawa couplings to $B_2$ tends to be larger than to $B_1$.
A similar plot for up-type VLQs is shown on the right where the coupling to $T_2$ also tends to be larger than the coupling to $T_1$.

\begin{figure}[h]
\includegraphics[scale=0.8]{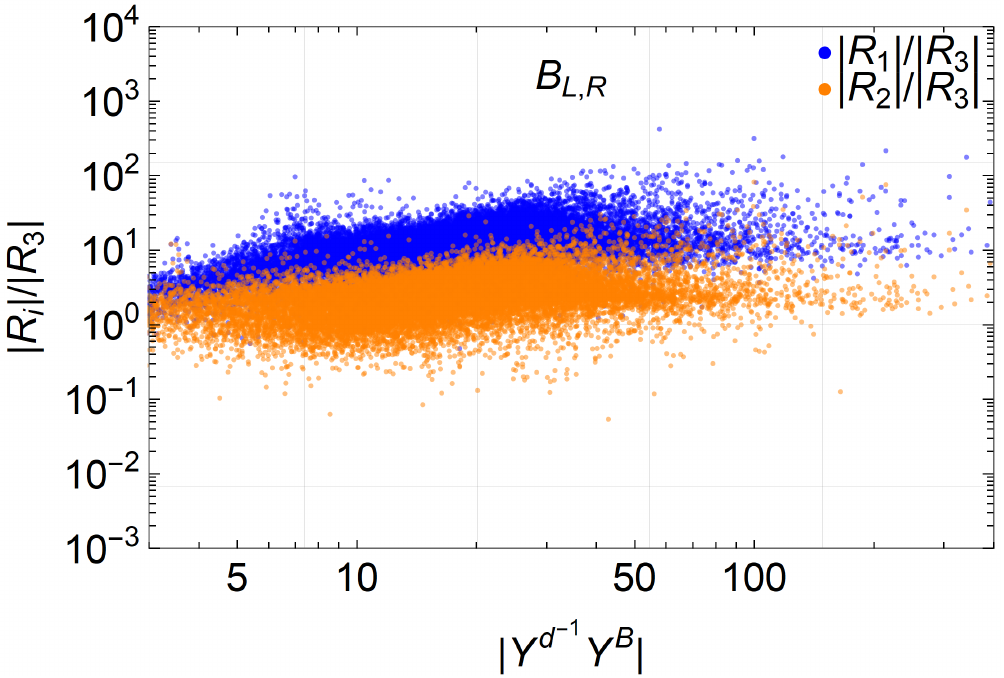}
\includegraphics[scale=0.8]{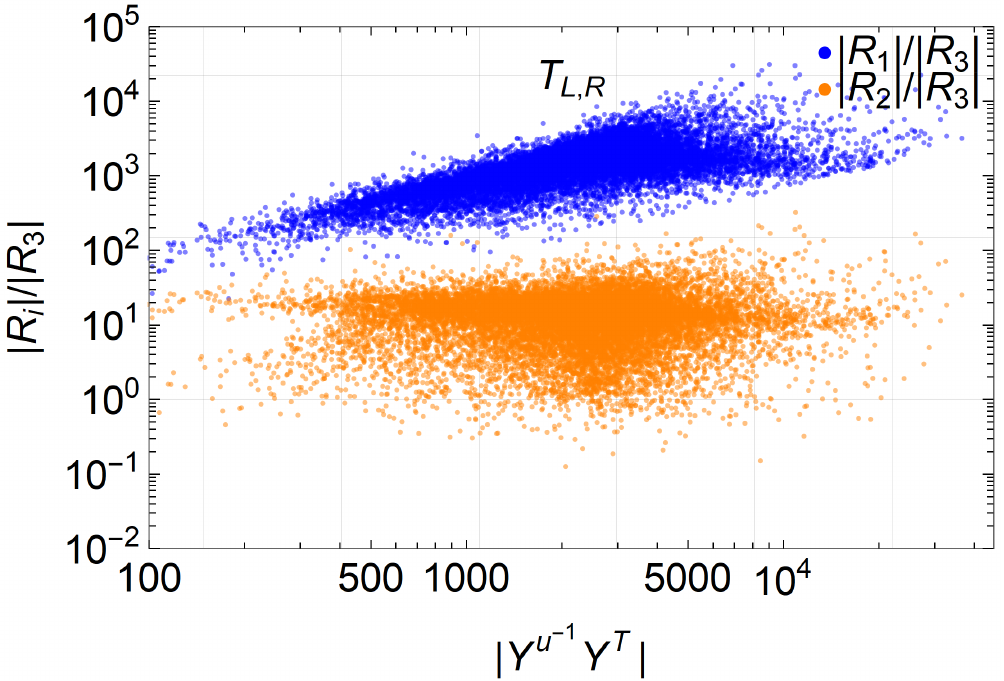}
\caption{\label{fig:ratios:n=2}
Ratios of $|R_i|/|R_3|$ in \eqref{Rdi:n=2} quantifying the strength of the Yukawa coupling to the NB-VLQs compared to the SM Yukawas.
}
\end{figure}

Finally, Fig.~\ref{fig:ratios:n=2} (left) shows the ratio between the different norms
\eq{
\label{Rdi:n=2}
|R^d_i|=\sqrt{|R^d_1|^2+|R^d_2|^2}
}
and $|R^d_3|$ as a function of the whole norm $|{Y^d}^{-1}Y^B|=|R^d|$.
The plot in the right is the same plot for up-type VLQs.
We can see that $|R_i|$ follow the same hierarchy for $n=1$ in \eqref{R:hierarchy}; see Fig.\,\ref{fig:ratios:n=1}.

\subsection{Inversion formula}
\label{sec:inversion:n=2}

In Sec.\,\ref{sec:param:n=1} we have reviewed the explicit parametrization devised in \cite{nb-vlq} for one NB-VLQ.
Such a parametrization allowed us to use the 10 SM flavor parameters as input and vary the additional five free parameters.
Here we want to devise a similar way to use the SM parameters as input and parametrize $\cY^d$ or $\cY^u$.

Let us focus on the down-type VLQ model, cf.\,\eqref{yuk:NB}, initially with an arbitrary number $n_B$ of VLQs.
Three of the 10 SM parameters are fixed in the diagonal $Y^u=\cY^u$.
Another 7 SM parameters should be accounted for in the down-sector including the VLQs. 
The task is then to solve for $\cY^d$ in \eqref{Yd:NB}, given the SM input in \eqref{Yd:sm.input}, with possible additional phases in \eqref{phases.beta}.

We first rewrite \eqref{Yd:NB} as 
\eq{
\label{inversion}
\id_3-ww^\dag = {\cY^d}^{-1}Y^d{Y^d}^{\dag}{\cY^d}^{-1\tp}\,. 
}
Then, after defining the real and imaginary parts,
\eqali{
Y^d{Y^d}^{\dag}&\equiv \La{}= \La1+i\La2\,,
\cr
\id_3-ww^\dag &\equiv \Omega=\Omega_1+i\Omega_2\,,
}
we separate the equation into the real and imaginary parts,
\subeqali{
\label{inversion:re}
\Omega_1 &= {\cY^d}^{-1}\La1{\cY^d}^{-1\tp}\,,
\\
\label{inversion:im}
\Omega_2 &= {\cY^d}^{-1}\La2{\cY^d}^{-1\tp}\,.
}
Note that $\La1,\La2$ are fixed from SM input, except for the phases $\beta_i$.

As $\Omega_1$ and $\La1$ are real symmetric and positive definite,%
\footnote{Because $\Omega$ is positive definite to satisfy \eqref{inversion}, then its real part $\Omega_1$ is also positive definite.}
the real part \eqref{inversion:re} can be solved for $\cY^d$ as
\eq{
\label{sol:calY}
\cY^d=\La1^{1/2}\cO \Omega_1^{-1/2}\,,
}
where $\cO$ is a real orthogonal matrix to be determined.
Plugging the form \eqref{sol:calY} into the imaginary part \eqref{inversion:im} leads to
\eq{
\label{inversion:im:O}
\Omega_1^{-1/2}\Omega_2\Omega_1^{-1/2}=\cO^\tp \La1^{-1/2}\La2\La1^{-1/2}\cO\,.
}
So $\cO$ is a matrix that transforms the known antisymmetric matrix in the righthand side, function of $\La{}=Y^d{Y^d}^\dag$, to another one depending on $w$ on the lefthand side.

Now we need to study how to parametrize $\Omega$ (or $ww^\dag$) considering that $\Omega$ is positive definite and $\id_3-\Omega=ww^\dag$ is positive semidefinite (ranks 1,2, and 3, for $n_B=1,2$, and $n_B\ge 3$).

We can try to simplify $\Omega$ by exploring the reparametrization freedom 
\eq{
\Omega\to O_{d_R}^\tp\Omega O_{d_R}\,,
}
which comes from 
\eq{
w\to O_{d_R}^\tp w\,,\quad
\cY^d\to \cY^dO_{d_R}\,,
}
induced by $d_R\to O_{d_R}d_R$. The matrix $O_{d_R}$ is real orthogonal.

We can go to the basis where
\eqali{
\label{ww:xy}
O_{d_R}^\dag(ww^\dag)O_{d_R}
    &= \mtrx{x_1&&\cr &x_2&\cr &&x_3}-i\mtrx{0&y_3&-y_2\cr -y_3&0&y_1\cr y_2&-y_1&0}\,,
}
where $x_{i}$, $y_{j}$ are real.
The eigenvalues $x_i$ of the real part are nonegative and we can choose the ordering $0\le x_1\le x_2\le x_3$.
Also, $x_i<1$ as $\Omega_1=\diag(1-x_i)$ is positive definite.
We avoid spurious cancellations and assume $\rank(ww^\dag)=\min(n_B,3)$.
For $n_B=1$, after a discrete choice, it is guaranteed that $x_1=0$, $y_2=y_3=0$ and $y_1^2=x_2x_3$; 
see appendix \ref{ap:A>=0}.
The parametrization of Sec.\,\ref{sec:param:n=1} corresponds to $x_2=b^2$ and $x_3=a^2$.
It is impossible that more than one $x_i$ vanishes for a complex matrix.

We can now analyze \eqref{inversion:im:O}, which can be rewritten as
\eq{
\label{y=b}
\ty_i=b_j\cO_{ji}\det\cO\,,
}
after defining the vectors $b_k$ and $\ty_k$ as
\eqali{
(\La1^{-1/2}\La2\La1^{-1/2})_{ij}&=\eps_{ijk}b_k\,,
\cr
(\Omega_1^{-1/2}\Omega_2\Omega_1^{-1/2})_{ij}&=\eps_{ijk}\ty_k\,.
}
In the basis \eqref{ww:xy}, we can find 
\eq{
\label{def:ty1}
\ty_1=\frac{y_1}{(1-x_2)^{1/2}(1-x_3)^{1/2}}\,,
}
and the other components follow from cyclic replacement: $1\to 2\to 3$.

Equation \eqref{y=b} tells us that the norm of $\ty_i$ should be equal to the norm of $b_i$ and $\cO$ should rotate one to the other.
Since the sign of $\det\cY^d$ can be chosen positive by eventually flipping $d_{R}\to -d_R$, the solution \eqref{sol:calY} allows us to choose $\det\cO=1$.
Let us denote by $\mu$ the norm:
\eq{
\mu\equiv \sqrt{b_ib_i}\,.
}
Note that this quantity is only a function of $\La{}$ and can be calculated as
\eq{
\mu=\sqrt{1-\frac{\det(\La{})}{\det(\re(\La1) )}}\,;
}
see appendix \ref{ap:A>=0}.
One can show that $0<\mu<1$\,\cite{nb-vlq}.
Then the equality of norms can be written as
\eq{
\label{sphere}
\ty_1^2+\ty_2^2+\ty_3^2=\mu^2\,.
}
Once $\ty_i$ is confined to this sphere, $\cO$ can be found from \eqref{y=b} as a rotation that connects it to $b_i$, except for rotations that leave $b_i$ invariant. 
We parametrize this freedom by an angle $\gamma$, which only affects \eqref{sol:calY}.

One difficulty is that not all the sphere in \eqref{sphere} is physical. We need to ensure $ww^\dag=\id_3-\Omega$ is positive semidefinite.
Let us first analyze the case where all $x_i\neq 0$. This covers the case $n_B\ge 3$ and most of $n_B=2$.
We prove in appendix \ref{ap:A>=0} that $\ty_i$ should be confined to the elipsoid
\eq{
\label{ellipsoid}
\frac{\ty_1^2(1-x_2)(1-x_3)}{x_2x_3}+
\frac{\ty_2^2(1-x_1)(1-x_3)}{x_1x_3}+
\frac{\ty_3^2(1-x_1)(1-x_2)}{x_1x_2}\le 1\,.
}
The inequality is for $n_B\ge 3$ and the equality is for $n_B=2$.
To have an intersection, the ellipsoid \eqref{ellipsoid} cannot be internal to the sphere \eqref{sphere} and then at least one of its semiaxes needs to be larger than $\mu$, so we need for the largest semiaxis,
\eq{
\label{largest.semiaxis}
\frac{x_2x_3}{(1-x_2)(1-x_3)}\ge \mu^2\,.
}
Note that the function $f(x)=x/(1-x)$ is a monotonically increasing function in the interval $[0,1)$ with $f(1/2)=1$.
For $x_2=x_3$, the equality is achieved for $x_2=\mu/(1+\mu)$ which is less than $1/2$.
If we introduce the shorthand notation
\eq{
f_i=f(x_i)\,,
}
we can rewrite \eqref{largest.semiaxis} as
\eq{
\label{f2.f3}
f_2f_3\ge \mu^2\,.
}
On the other extreme, if the smallest semiaxis obeys
\eq{
f_1f_2\ge \mu^2\,,
}
then the whole sphere is allowed.

If $x_1=0$ ($\ty_2=\ty_3=0$), then \eqref{sphere} implies that $\ty_1^2=\mu^2$.
As $\Omega$ becomes block diagonal, the condition \eqref{ellipsoid} is replaced exactly by \eqref{f2.f3} where the inequality is valid for $n_B=2$ and the equality is for $n_B=1$.
The case of inequality restricts the region for $(x_2,x_3)$ in the unit square $(0,1)\times(0,1)$, restricted to $x_2<x_3$.

Let us now specialize to $n_B=2$.
The subcase $x_1=0$ was treated above so we assume $x_1>0$.
If $x_1>0$, we can isolate $f_3$ in \eqref{ellipsoid} for the equality as
\eq{
\label{f3}
f_3=\frac{1}{f_1f_2-\ty_3^2}\left(f_1\ty_1^2+f_2\ty_2^2\right)\,,
}
which is a solution once $f_3>0$ is ensured.
Once $\ty_i$ is restricted to the sphere \eqref{sphere}, 
condition \eqref{largest.semiaxis} is indeed a necessity. Therefore we need to know the values of $f_1$, $f_2$ and all $\tilde{y}_i $ to determine $f_3$. In the case of $f_1 $ and $f_2$, we just need to know the values of $x_1$ and $x_2$.
As $\tilde{y}_i$ are restricted to the sphere \eqref{sphere}, we parametrize
\begin{equation}\label{tildeymuphi}
    \begin{split}
    \tilde{y}_1&=\mu\sin(\phi_2)\cos(\phi_1),\\
      \tilde{y}_2&=\mu\sin(\phi_2)\sin(\phi_1),\\
     \tilde{y}_3  &=\mu \cos(\phi_2).
    \end{split}
\end{equation}
Thus, $x_3$ is determined from $f_3$ which in turn is a function of $\La{}$ and the parameters
$\{x_1,x_2,\phi_1,\phi_2\}$.

Once $x_3$ is determined, $\Omega_1$ is obtained from 
\begin{equation}
\Omega_1=\diag(1-x_1,1-x_2,1-x_3)\,.   
\end{equation}
Finally, $\cY^d$ is determined from the inversion formula \eqref{sol:calY} which depends on the SM parameters in $\La{}$ and additionally on 
\begin{equation}
\label{free:n=2:calY}
    \{x_1,x_2,\phi_1,\phi_2,\gamma,\beta_1,\beta_2\}.
\end{equation}
For the case $x_1=0$, the additional parameters of the inversion formula becomes 
\eq{
\{x_2,x_3,\gamma,\beta_1,\beta_2\}\,.
}

Let us check the number of parameters so far for 
$n_B=2$.
In this case, there are 21 parameters in total among which 10 must account for the SM flavor parameters.
Three of the latter are just the three up-type quark masses of the SM.
Another 7 SM parameters reside in $\La{}$. The remaining 11 parameters should be free, 7 of which are listed in \eqref{free:n=2:calY}.
There are still 4 more parameters that are not needed for $\cY^d$.
We treat them in the following.

\subsection{Additional parameters in $w$ and $\cM^B$}
\label{sec:n=2:w.cMB}

In Sec.\,\ref{sec:inversion:n=2}, to account for the SM flavor parameters in $Y^d{Y^d}^\dag$ to describe $\cY^d$ -- the inversion formula --, we have chosen the basis \eqref{ww:xy} for $ww^\dag$.
However, additional parameters are still needed to describe $w$ itself and this  quantity enters in $Y^B$ in \eqref{YB:cal-Yd}. The rest of parameters resides in $\cM^B$.

We first analyze $w$. Being a complex matrix of size $3\times n_B$, we can write  
the singular value decomposition of $w$ as
\eq{
\label{w:svd}
w=U_w\hw V_w^\dag\,,
}
where $U_w$ and $V_w$ are $3\times 3$ and $n_B\times n_B$ unitary matrices respectively.
The matrix $U_w$ and the singular values in $\hw$ are fully determined from the parameters $x_i,y_i$ in \eqref{ww:xy}, except for rephasing of columns of $U_w$.
This freedom will be absorbed in $V_w^\dag$. Then the parameters in $V_w^\dag$ are the additional parameters in $w$.

Specializing now to $n_B=2$, we use
\begin{equation}\label{w:sv:n=2}
\hw= \begin{pmatrix}
        0&0\\
        \hw_1&0\\
        0&\hw_2\\
    \end{pmatrix}\,.
\end{equation}
In the basis \eqref{yuk:VLQ}, rephasing of the fields $B_{aR}$ leads to rephasing from the right to $w$ and then to $V_w^\dag$.
So $V_w^\dag$ is a $2\times 2$ unitary matrix with rephasing freedom from the right and we can parametrize it using two additional parameters:
\begin{equation}
\label{param:Vw:n=2}
V_w^\dag=\begin{pmatrix}
        e^{i\psi_2}\cos(\psi_1)&  \sin(\psi_1)\\
         -\sin(\psi_1)& e^{-i\psi_2}\cos(\psi_1)\\
    \end{pmatrix}.
\end{equation}

We can now analyze $\cM^B\sim n_B\times n_B$. In the original basis \eqref{yuk:NB}, there is still freedom that allows us to choose diagonal
\eq{
\label{diag:calMB}
\cM^B=\hat{\cM}^B=\diag(\mu_1,\mu_2,\dots,\mu_{n_B})\,.
}
These parameters complete the number of parameters and the mass matrix for the heavy VLQs are determined from  \eqref{WR:MB}. Note that the latter  is non-diagonal and complex in general.
For $n_B=2$, we have two free mass parameters $\mu_1,\mu_2$.

Hence, for $n_B=2$, the four additional parameters that are unnecessary for $\cY^d$ are
\eq{
\label{n=2:additional.param}
\{\psi_1,\psi_2,\mu_1,\mu_2\}\,.
}
These 4 parameters, together with the 7 in eq.\,\eqref{free:n=2:calY} that enters in $\cY^d$, complete the 11 free parameters besides the SM parameters.
A similar analysis can be extended to general $n_B$.

\subsection{Heavy mass matrix}

In the parametrization described in Secs.\,\ref{sec:inversion:n=2} and \ref{sec:n=2:w.cMB}, besides the SM parameters present in $Y^d{Y^d}^\dag$, 
we use $w$ and diagonal $\cM^B$ as input.
These quantities specify the various parameters in the original Lagrangian \eqref{yuk:NB}: 
$\cY^d$ follows from the inversion formula \eqref{sol:calY} while $\cM^{Bd}$ follows from \eqref{def:w} or \eqref{wdagger} as 
\eq{
\cM^{Bd}=\cM^B\tw^\dag=
M^B w^\dag\,.
}
The mass matrix $M^B$ for the heavy VLQs is given by  \eqref{WR:MB}. Note that it is non-diagonal for diagonal $\cM^B$.
Here we analyze the various relations between $M^B$ and the other parameters.

Specifically for $n_B=2$, the parameters \eqref{free:n=2:calY} are necessary to parametrize $\cY^d$ and $ww^\dag$, while \eqref{n=2:additional.param} are additionally necessary to specify $w$ and $\cM^B$.
We continue the discussion below restricted to $n_B=2$.

We can first parametrize the singular values in \eqref{w:sv:n=2} in terms of angles $\theta_i$:
\eq{
\hw_i=s_i\equiv \sin\theta_i\,,
}
with $\theta_i\in [0,\pi/2]$.
Typically these singular values are close to unity and depend on the parameters $x_i,y_i$ in \eqref{ww:xy}.
In terms of $\theta_i$, the matrix $\tw$ in \eqref{def:w} or \eqref{YB:w-tilde} is
\eq{
\label{wtilde:tan}
\tw={Y^d}^{-1}Y^B=U_w
\begin{pmatrix}
        0&0\\
        t_1 &0\\
        0& t_2\\
    \end{pmatrix}V_w^\dag\,,
}
where $t_i\equiv\tan\theta_i$ and the matrices $U_w,V_w$ are the same as in \eqref{w:svd}.
Typically $t_i\gg 1$.
Note that its norm is the same as the norm of $R^d$, cf.\,\eqref{R.norm}.

In the same way, the mass matrix \eqref{WR:MB} for the heavy VLQs becomes
\eq{
M^B=\hat{\cM}^B V_w \diag\Big(\frac{1}{c_1},\frac{1}{c_2}\Big)V_w^\dag\,,
}
where $c_i\equiv \cos\theta_i$ and $V_w$ is parametrized as \eqref{param:Vw:n=2}.
For fixed values of $\mu_i$ in $\hat{\cM}^B$, 
we note that large values of $t_i$ in 
\eqref{wtilde:tan} leads to large values of $1/c_i$ in $M^B$ and it tends to lead to larger VLQ masses.

Instead of $\mu_1,\mu_2$, we could equally use $\mu_0,\kappa$ as
\eq{
\label{MB:kappa}
M^B=\mu_0\diag(\kappa,\kappa^{-1})V_w\diag\Big (\frac{1}{c_1},\frac{1}{c_2}\Big)V_w^\dag\,.
}
Using this form we show in Fig.\,\ref{fig:MB:n=2:k=1,1-100} (left) the singular values for $M^B/\mu_0$ for down-type VLQs. The analogous plot for up-type VLQs are shown on the right.
The darker points are for fixed $\kappa=1$ while the lighter points are obtained by varying $\kappa\in[1,100]$.
For the latter, choosing $\kappa\in [0.01,1]$ instead leads to a similar result.
For $\kappa=1$, we can see that the lighter mass is approximately given by $\mu_0$ while the larger mass increases with $|R^d|$ or $|R^u|$.
Relative to $\mu_0$, the points in the plots also correspond to $1/c_1,1/c_2$ in \eqref{MB:kappa}.
Note the mass hierarchy for the up-type VLQs is stronger.
Incidentally, $\kappa=1$ corresponds to the CP4 model proposed in Ref.\,\cite{NB:CP4}.
As $\kappa$ is allowed to vary, we can see that the masses get scattered around the values for $\kappa=1$ with relative variation of the order of the maximum $\kappa$.
\begin{figure}[h]
\includegraphics[scale=0.8]{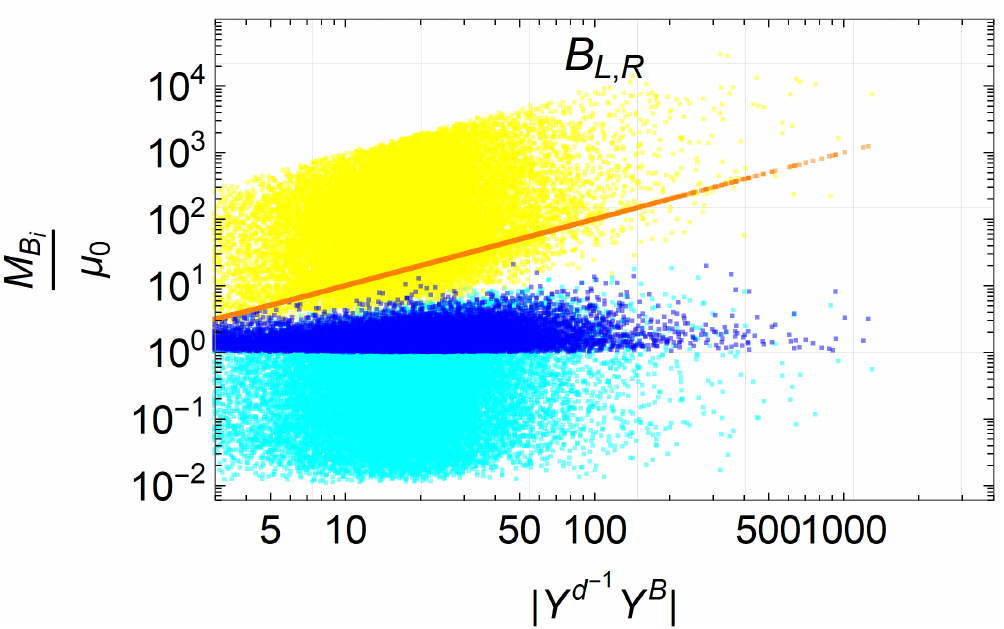}
\includegraphics[scale=0.8]{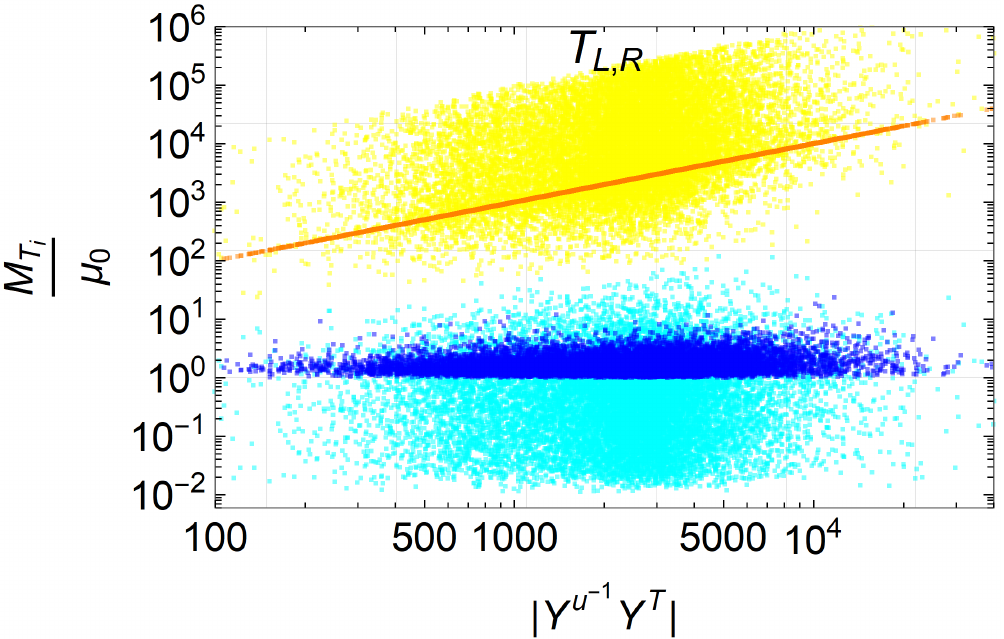}
\caption{\label{fig:MB:n=2:k=1,1-100}%
VLQ masses $M_{Q_1},M_{Q_2}$ relative to $\mu_0$ in \eqref{MB:kappa} against $|(Y^q)^{-1}Y^Q|$ for NB-VLQ of down ($q=d,Q=B$) and up ($q=u,Q=T$) types.
The darker colors (dark blue and orange) refer to $\kappa=1$ and the lighter colors (cyan and yellow) refer to $\kappa\in [1,100]$.
}
\end{figure}

\subsection{Invariants for $\btheta$ for $n=2$}

The invariants shown in \eqref{n=1:invariants},
for one VLQ, only involved two insertions of the VLQ Yukawa $Y^Q$, $Q=B,T$.
With two or more families of VLQs, one can construct invariants involving more insertions of $Y^Q$ and insertions of the SM up quark Yukawa $Y^u$ but without the down quark Yukawa $Y^d$\,\cite{vecchi.1}.
In principle, these invariants will lead to larger estimates of $\btheta$, hence to more stringent constraints.

Considering two or more NB-VLQs of either down-type or up-type, Ref.\,\cite{vecchi.1} finds the following invariants as the dominant ones \adriano{for} estimating $\btheta$:
\eqali{
\label{invariants:n=2}
\text{down-type $n_B\ge 2$: }&\quad
\left(\frac{1}{16 \pi^2}\right)^3\im \Tr\left( \left[{Y^B}^\dag Y^u{Y^{u}}^\dagger {Y^B}, {Y^B}^\dag{Y^B}\right]F\left({M^B}^\dagger M^B\right)\right),
\cr
\text{up-type $n_T\ge 2$: }&\quad
\left(\frac{1}{16 \pi^2}\right)^3\im \Tr\left( \left[\tY^{T^{\mfn{\dag}}} Y^u{Y^{u}}^\dagger \tilde{Y}^T, \tY^{T^{\mfn{\dag}}}\tilde{Y}^T\right]F\left({M^T}^\dag M^T\right)\right),
}
where we use for the function $F$ the form
\begin{equation}
    F\left({M^Q}^\dagger M^Q\right)=\frac{{M^Q}^\dagger M^Q}{\Tr\left({M^Q}^\dagger M^Q\right)}\,,
\end{equation}
for $Q=B,T$.
Note that for both invariants in \eqref{invariants:n=2}, we are adopting the basis where $Y^u=\hY^u$ is diagonal and then for the up-type case, the VLQ Yukawa should be $\tY^T$; cf.\,\eqref{tildeYB}.

The invariant above can be further estimated in terms of SM Yukawas and $|R^d|$ or $|R^u|$.
These estimates can be found in Ref.\,\cite{vecchi.1} but, similarly to the $n=1$ case, the hierarchy of the components of $|R_{ia}|$ were not correct.
Taking the correct typical hierarchy shown in Fig.\,\ref{fig:ratios:n=2}, we can correct the estimate in \eqref{invariants:n=2} for two down-type VLQs as 
\begin{equation}
\label{inv:n=2:estimate:down}
\bar{\theta}\sim  \frac{\lambda^2_Cy^3_b y^2_ty_s}{\left(16 \pi^2\right)^3}\left(\frac{|R^d|}{10\sqrt{2}}\right)^4\sim 6\times 10^{-18} \left(\frac{|R^d|}{10\sqrt{2}}\right)^4
\,.
\end{equation}
Similarly, the invariant for two up-type VLQs in \eqref{invariants:n=2} can be estimated as 
\begin{equation}
\label{inv:n=2:estimate:up}
\bar{\theta}\sim  \frac{y^4_ty^2_c}{\left(16 \pi^2\right)^3}\left(\frac{|R^u|}{10^3\sqrt{2}}\right)^4\sim 10^{-12}\left(\frac{|R^u|}{10^3\sqrt{2}}\right)^4
\,.
\end{equation}
Here the factor $\sqrt{2}$ is specific for $n=2$.

\begin{figure}[h]
\includegraphics[scale=0.8]{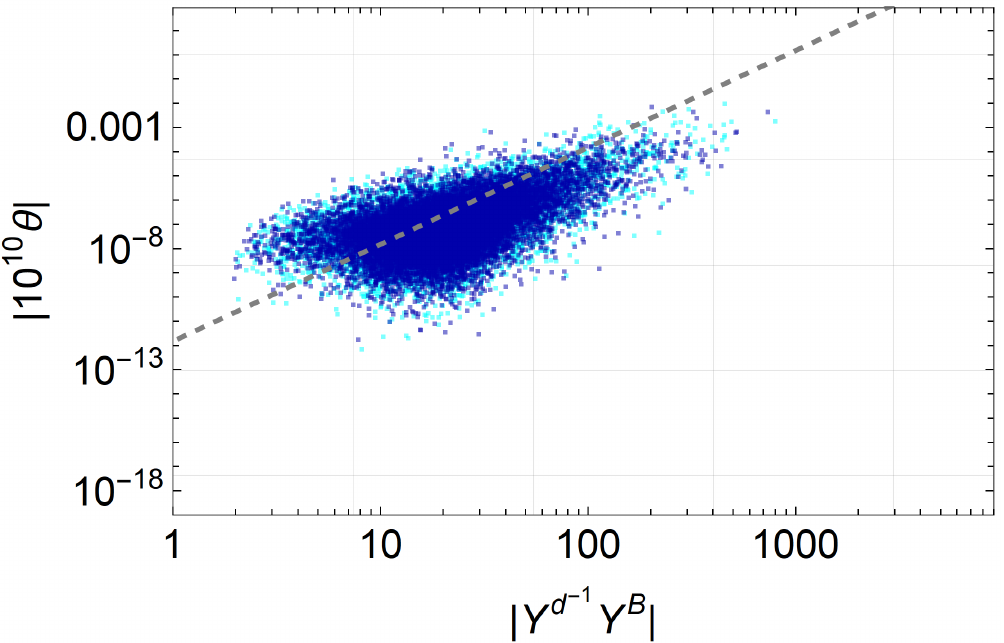}
\includegraphics[scale=0.8]{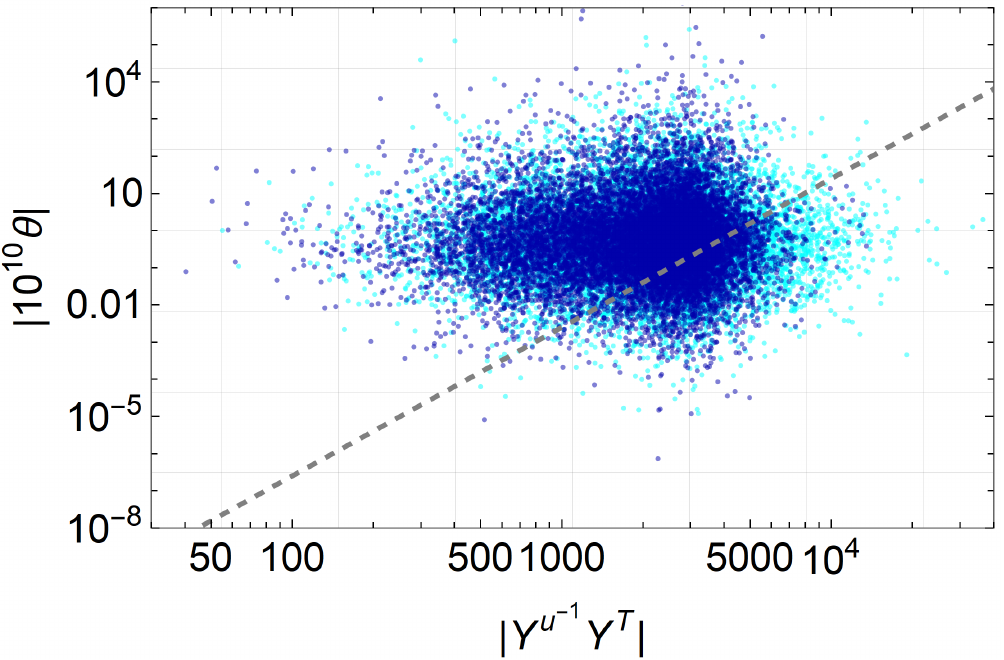}
\caption{\label{fig:theta:n=2}%
Invariants \eqref{invariants:n=2} estimating 3-loop contributions to $\bar{\theta}$ as a function of $|(Y^q)^{-1}Y^Q|=|R^q|$ for NB-VLQ of down type ($q=d,Q=B$) in the left plot and up type ($q=u,Q=T$) in the right plot.
Dark blue points denote $\mu_0=1.2\,\unit{TeV}$ ($\mu_0=1.3\,\unit{TeV}$) while light blue points denote $\mu_0=12\,\unit{TeV}$ ($\mu_0=13\,\unit{TeV}$) for the down type (up type). We fix $\kappa=1$.
Dashed lines denote the estimates \eqref{inv:n=2:estimate:down} and \eqref{inv:n=2:estimate:up}.
}
\end{figure}

We show in Fig.\,\ref{fig:theta:n=2} (left) the invariant of down-type in \eqref{invariants:n=2} as a function of $|R^d|$.
A similar plot for up-type VLQs is on the right.
For the masses, we fix $\kappa=1$ and choose two values for $\mu_0$, one around 1 TeV (dark blue) and the other one around 10 TeV (light blue).
We filter through flavor and electroweak observables discussed in Sec.\,\ref{sec:flavor:n=2}. 
We anticipate that their constraints are weaker for higher masses, allowing to estimate their impact by comparing the light blue and dark blue regions.
The estimate functions in \eqref{inv:n=2:estimate:down} and \eqref{inv:n=2:estimate:up} are shown in dashed lines. 
The correctness of these estimates can be checked as the lines pass through the middle of the points.
However, the scattered points show that a large deviation from the estimate is possible, in some cases of a few orders of magnitude.

We can see that for down-type VLQs, the constraint on $\btheta$ is very weak. The flavor and electroweak constraints (difference between light blue and dark blue) do not lead to visible differences by changing $\mu_0=1.2\,\unit{TeV}$ to $\mu_0=12\,\unit{TeV}$.
In contrast, for up-type NB VLQs, $\btheta<10^{-10}$ roughly constrains half the points. 
Even if we consider that the invariant \eqref{invariants:n=2} contributes to $\btheta$ with a prefactor of order 10, there is still plenty of points that pass the constraints.
As for the flavor and electroweak constraints, we can see that more points are allowed for the higher masses. 
Finally, the special points that led to a large suppression on $\btheta$ for $n=1$, cf. Fig.\,\ref{fig:theta:n=1}, are not present for $n=2$ and no significant suppression is seen if we choose one column or row of CKM real.
So we do not show these points separately.

\subsection{Flavor and electroweak constraints}
\label{sec:flavor:n=2}

For one VLQ of down-type that couples exclusively with the third family, one of the strongest contraints is given by the measurement of $R_b\equiv \Gamma(Z\to b\bar{b})/\Gamma(Z\to \text{hadrons})$\,\cite{saavedra:handbook}.
This constraint can still be applied with more than one VLQ.
In terms of $X_{bb}$ which enters in the neutral current to the $Z$, we obtain the bound
\begin{equation}
\label{eq:Rb-constraint}
    X_{bb}-0.2815(X_{dd}+X_{ss})=0.4381\pm0.0017.
\end{equation}
At leading order, we can write
$
X_{ij}\approx \delta_{ij}-(\Theta\Theta^{\dag})_{ij}
$,
where $\Theta$ is defined in \eqref{def:Theta}.
As can be seen in Fig.\,\ref{fig:YQ:n=2}, the VLQ Yukawa couplings to the third family are naturally stronger and they are larger for larger $|R^d|$.
In the region of large $|R^d|$, for $\kappa=1$, Fig.\,\ref{fig:MB:n=2:k=1,1-100} tells us that the hierarchy of masses overcomes the mild hierarchy of $Y^B_{3a}$ in Fig.\,\ref{fig:Y31Y32:n=2}, and we can assume $|\Theta_{31}|$ dominantes in $X_{bb}$.
For varying $\kappa$, the dominance of $\Theta_{31}$ is still valid.
Therefore, \eqref{eq:Rb-constraint} leads, at $2\sigma$, to the constraint
\eq{
|\Theta_{31}|<0.05\,.
}
This corresponds to $|\tY^B_{31}|<0.33$ for $M_{B_1}=1.2\,\unit{TeV}$.
This is never achieved for $|Y^B_{31}|\approx |\tY^B_{31}|$ in Fig.\,\ref{fig:Y31Y32:n=2} and this constraint is satisfied for all points.

For an up-type VLQ coupling exclusively with the third family, the strongest constraint comes from the oblique parameters $S,T$\,\cite{saavedra:handbook}.
We can extend this constraint to two up-type VLQs by using the expressions\,\cite{Lavoura:1992np}:
\begin{equation}
\begin{split}
\Delta S&=\frac{3}{2\pi}\Bigg[
    (|V_{tb}|^{2}-1)\psi(y_t,y_b)
    +|V_{T_1 b}|^{2}\psi(y_{T_{1}},y_b)
    +|V_{T_2 b}|^{2}\psi(y_{T_{2}},y_b)
    \cr&\hspace{3em}
    -\ |X_{T_1 t}|^{2}\chi(y_{T_1},y_t)
    -|X_{T_2 t}|^{2}\chi(y_{T_2},y_t)\Bigg],
\\
\Delta T&=\frac{3}{16\pi s^2_Wc^2_W}\Bigg[ 
    (|V_{tb}|^{2}-1)\theta(y_t,y_b)
    +|V_{T_1 b}|^{2}\theta(y_{T_1},y_b)
    +|V_{T_2 b}|^{2}\theta(y_{T_2},y_b)
    \cr&\hspace{7em}
    -|X_{T_1 t}|^{2}\theta(y_{T_1},y_t)
    -|X_{T_2 t}|^{2}\theta(y_{T_2},y_t)\Bigg],
\end{split}\end{equation}
with the loop functions given by
\begin{equation}\begin{split}
    \chi(y_1,y_2)&=\frac{5(y^2_1+y^2_2)-22y_1y_2}{9(y_1-y_2)^2}+ \frac{3y_1y_2(y_1+y_2)-(y^3_1+y^3_2)}{3(y_1-y_2)^3}\log\frac{y_1}{y_2},\\
    \theta(y_1,y_2)&=(y_1+y_2)-\frac{2y_1y_2}{y_1-y_2}\log\frac{y_1}{y_2},\\
    \psi(y_\alpha,y_i)&=\frac{1}{3}-\frac{1}{9}\log\frac{y_\alpha}{y_i}\,,
\end{split}\end{equation}
for $y_i\equiv m^2_i/m^2_Z$.

Recently, the CDF collaboration released a new analysis for the $W$-boson measurement\,\cite{CDF:2022hxs}, which may significantly modify the values of the oblique parameters S, T (hereafter we only consider the case  $\Delta U=0$)\,\cite{deBlas:2022hdk,Lu:2022bgw}.  
Since the main focus of our work is on characterising NB-VLQs, we will conservatively consider the values for S,T obtained from an electroweak global fit previous to the CDF new analysis\,\cite{ParticleDataGroup:2020ssz}
\begin{equation}\label{eq:WPDG}
    \Delta S = 0.00 \pm 0.07, \quad \Delta T = 0.05 \pm 0.06, \quad \text{correlation = } 0.92\,.
\end{equation}
Using \eqref{eq:WPDG} at 95\% CL, we require
\eq{
\Delta\chi^2=\frac{1}{1-\rho^2}(\bar{x}^2+\bar{y}^2-2\rho\bar{x}\bar{y})
<5.99146\,,
}
where $\rho$ is the correlation, $\bar{x}=(x-\mu_x)/\sigma_x$ for $x=\Delta S$ and, similarly, $y=\Delta T$.
We show as red points in Fig.\,\ref{fig:YQ:n=2} the points that do not pass this constraint.
We can see that the Yukawas that pass the contraint obey
\eq{
\sqrt{|Y^T_{31}|^2+|Y^T_{32}|^2}\lesssim 5\times \frac{1.3\,\unit{TeV}}{{M_{T_1}}}
\,,
}
and larger Yukawas start to be cut.
For these points, we are using $\kappa=1$ and $\mu_0=1.3\,\unit{TeV}$, which roughly corresponds to the lightest mass $M_{T_1}$; cf.\,Fig.\,\ref{fig:MB:n=2:k=1,1-100}.

To constrain less frequent but possible VLQ Yukawas away from the typical hierarchy \eqref{typical.YB} or \eqref{typical.YT}, we should also consider flavor changing processes involving the first and second families.
For simplicity, we focus on $\Delta F=2$ transitions which are the strongest.
For large VLQ mass, the effective Lagrangian for neutral meson mixing, coming from box diagrams with Higgs exchange, can be written as\,\cite{IshiwataLigetiWise2015} 
\eq{
\lag_{\text{meson}}=-\frac{\Lambda_{ij}}{128\pi^{2}}\left[\sum_{klmn}(\bar{u}_{L}^{k}V_{ki}\gamma_{\mu}V^{\dag}_{jl}u_{L}^{l})(\bar{u}_{L}^{m}V_{mi}\gamma^{\mu}V^{\dag}_{jn}u_{L}^{n})+(\bar{d}_{L}^{i}\gamma_{\mu}d_{L}^{j})(\bar{d}_{L}^{i}\gamma^{\mu}d_{L}^{j})\right]\,,
}
where 
\eq{
\Lambda_{ij}=\frac{1}{M^2_Q}\left(
    Y^Q_{i1}Y^{Q*}_{j1}+Y^Q_{i2}Y^{Q*}_{j2}
\right)^2\,,
}
for the VLQ $Q=B,T$ exchange.
For simplicity, we assumed the common mass $M^2_Q$ for the VLQs.
Also note that $Y^Q\to \tY^B$ for down-type VLQs.
Now, the constraints obtained in Ref.\,\cite{IshiwataLigetiWise2015} for two up-type VLQs read
\eqali{
|Y_{31}^{T^*} Y_{11}^T+Y_{32}^{T^*} Y_{12}^T|\leq \frac{M_T}{25\;\unit{TeV}},
\cr
|Y_{31}^{T^*} Y_{21}^T+Y_{32}^{T^*} Y_{22}^T|\leq \frac{M_T}{6.4\;\unit{TeV}},
\cr
|\text{Re}(Y_{11}^{T^*} Y_{21}^T+Y_{12}^{T^*} Y_{22}^T)|\leq \frac{M_T}{42\;\unit{TeV}},
\cr
|\text{Im}(Y_{11}^{T^*} Y_{21}^T+Y_{12}^{T^*} Y_{22}^T)|\leq \frac{M_T}{670\;\unit{TeV}}\,.
}
For two down-type VLQs, the same is valid with the exchange $M_T\to M_B$ and $Y^T_{ia}\to \tY^B_{ia}$.

Only points that satisfy all the constraints of this subsection are shown in Fig.\,\ref{fig:theta:n=2}.
We can see that for up-type VLQs more points extends on the right for the case of the larger mass (light blue) while for down-type VLQs there is no appreciable difference.

\section{Summary}
\label{sec:summary}

Generalizing the parametrization found in Ref.\,\cite{nb-vlq} for a single VLQ of Nelson-Barr type of either down or up-type, we have proposed an explicit parametrization for two singlet NB-VLQs.
With the explicit parametrization in hand, we studied several aspects of the model.
In special, we analyzed the flavor invariants presented in Ref.\,\cite{vecchi.1} which estimate the 3-loop contribution to $\btheta$ that arises solely from the VLQ Yukawas needed for the transmission of CP violation to the SM.
As shown in Fig.\,\ref{fig:theta:n=2}, for two down-type NB-VLQs, the constraint arising from $\btheta$ is very weak.
On the other hand, confirming Ref.\,\cite{vecchi.1}, the constraint on two up-type NB-VLQs is very strong.
Here, using the explicit parametrization, we were able to be more quantitative and found out that the $\btheta$ constraint eliminates roughly half the points if the estimate is taken with coefficient unity. 
But even if the $\btheta$ invariant underestimates the real contribution by two orders of magnitude, there is still room for TeV scale NB-VLQs.
Therefore, it is crucial that an explicit parametrization be used for a detailed and quantitative analysis, as some quantities may deviate from the estimates by few orders of magnitude.

We briefly summarize other important points that were analyzed in the paper:
\begin{itemize}
\item  We have reviewed the case of a single NB-VLQ ($n=1$) and have explicitly shown the distribution of the VLQ Yukawas in Fig.\,\ref{fig:YQi:n=1}, illustrating their hierarchy: the NB-VLQ couples more strongly with the heavier SM quark.

\item For $n=1$, we corrected the numerical estimates, found in Ref.\,\cite{vecchi.1}, of the flavor invariants for $\btheta$.
The scatter plot for $\btheta$ is shown in Fig.\,\ref{fig:theta:n=1}.
The correction is mainly due to a better estimate of the hierarchy for $R_i^q$, $q=u,d$; see Fig.\,\ref{fig:ratios:n=2}.
Another aspect is that an explicit parametrization gives the full description which cannot be captured by simple estimates.
For this reason the range of possible $|R^q|$ is also seen to be larger, cf.\,\eqref{range:R}:
\eqali{\nonumber
2&\lesssim |R^d|\lesssim 10^4\,,
\cr
30&\lesssim |R^u|\lesssim 3\times 10^5\,.
}
In the Nelson-Barr setting, they quantify the ratio between the CP violating contribution and the CP conserving bare mass of the VLQs.
They are expected to be larger than unity but we see they span quite a range.

\item We also showed that for $n=1$, there are special points (cf.\ Sec.\,\ref{sec:special}) where the VLQ Yukawas to the third SM family vanish, making the $\btheta$ invariants extremely suppressed; see
Fig.\,\ref{fig:theta:n=1}.
These special points were analyzed first for the case of vector-like leptons transmitting CP violation\,\cite{nb-vll}.

\item  For $n=2$, we show explicitly the distributions of various quantities.
In particular, Fig.\,\ref{fig:YQ:n=2} shows the norm of the VLQ Yukawas that couple to each SM family. They obey the same approximate hierarchy of the $n=1$ case, cf.\,\eqref{typical.YB} and \eqref{typical.YT}.

\end{itemize}

\acknowledgments

G.H.S.A.\ acknowledges financial support by the Coordenação de Aperfeiçoamento de Pessoal de Nível Superior - Brasil (CAPES) - Finance Code 001.  A.C.~acknowledges support from National Council for Scientific and Technological Development – CNPq through projects 166523\slash2020-8 and 201013\slash2022-3.
C.C.N.\ acknowledges partial support by Brazilian Fapesp, grant 2014/19164-6, and CNPq, grant 312866/2022-4.

\appendix
\section{Partial diagonalization}
\label{ap:partial}

The changing of basis from \eqref{yuk:NB} to \eqref{yuk:VLQ} is easily described by comparing the complete mass matrix of the down-type quarks following from each case after EWSB:
\eq{
\label{mass.matrix}
\text{NB:}\quad\cM^{d+B}=\mtrx{\frac{v}{\sqrt{2}}\cY^d & 0\cr \cM^{Bd}  & \cM_B}
\,,\quad
\text{generic:}\quad M^{d+B}=\mtrx{\frac{v}{\sqrt{2}}Y^d & \frac{v}{\sqrt{2}}Y^B\cr 0 & M_B}\,.
}
Only a unitary transformation from the right is necessary to connect them:
\eq{
\label{basis:real.gen}
\cM^{d+B}W_R=M^{d+B}\,.
}

We can find an explicit form for $W_R$ in \eqref{basis:real.gen} by assuming
\eq{
\label{WR:w}
W_R=\mtrx{\big(\id_3-ww^\dag\big)^{1/2} & w\cr -w^\dag & \big(\id_n-w^\dag w\big)^{1/2}}\,.
}
It will be easier, however, to use a slightly different parametrization
\eq{
W_R=\mtrx{\id_3 & \tw \cr -\tw^\dag & \id_n}
\mtrx{(\id_3+\tw\tw^\dag)^{-1/2} & 0 \cr 0 & (\id_n+\tw^\dag \tw)^{-1/2}}
\,,
}
valid for $\tw\neq 0$, where both matrices are not unitary but the product is.
The crucial point is that only the first matrix matters to guarantee the zero block in $M^{d+B}$ so that
the solution is easily 
\eq{
\label{w-tilde}
\tw^\dag = {\cM^B}^{-1}\cM^{Bd}\,,
}
where $\cM^B$ is guaranteed to be nonsingular.
We can write everything in terms of $w$ if needed.
For example,
\eq{
\label{tw:w}
w=\tw(\id_n+\tw^\dag\tw)^{-1/2}
=(\id_3+\tw\tw^\dag)^{-1/2}\tw\,.
}
If $W_R\sim 2\times 2$ were real, we would have $w= \sin\theta$ while $\tw=\tan\theta$ for some angle $\theta$.

Then we get explicitly
\subeqali{
Y^d&= \cY^d(\id_3+\tw\tw^\dag)^{-1/2}=\cY^d(\id_3-ww^\dag)^{+1/2}\,,
\\
Y^B&= \cY^d \tw(\id_n+\tw^\dag \tw)^{-1/2}=\cY^dw\,,
\\
M^B&=\cM^B(\id_n+\tw^\dag \tw)^{+1/2}=\cM^B(\id_n-w^\dag w)^{-1/2}\,.
}
These are the relations in \eqref{WR:YdYB}.

\section{Parameter region for positive (semi)definite matrix}
\label{ap:A>=0}

Let $A\sim 3\times 3$ be a complex positive definite or positive semidefinite matrix.
So this discussion applies to $Y^d{Y^d}^\dag$, to $ww^\dag$ or to $\Omega=\id_3-ww^\dag$.

We first decompose $A$ into real and imaginary parts:
\eq{
A=A_1+iA_2\,.
}
Allowing real orthogonal basis change, we choose a specific basis where $A_1$ is diagonal:
\eq{
\label{basis:diag.A1}
A_1=\diag(a_1,a_2,a_3)\,,\quad
(A_2)_{ij}=\eps_{ijk}b_k\,.
}
Note that $A_1$ is positive (semi)definite if $A$ is positive (semi)definite.
Then $a_i\ge 0$ and we can order $a_1\le a_2\le a_3$.
We assume $b_ib_i\neq 0$ to exclude real $A$.

If we calculate the characteristic equation,
\eq{
\det(A-\lambda I)=-\big[\lambda^3-\gamma_1(A)\lambda^2-\gamma_2(A)\lambda-\gamma_3(A)\big]\,,
}
we obtain in the basis \eqref{basis:diag.A1},
\eqali{
\label{gammai:A}
\gamma_1(A)&=\Tr[A]=a_1+a_2+a_3\,,\cr
\gamma_2(A)&=\ums{2}\Tr[A^2-\gamma_1(A)A]=b_1^2+b_2^2+b_3^2-a_1a_2-a_2a_3-a_3a_1\,,\cr
\gamma_3(A)&=\ums{3}\Tr[A^3-\gamma_1(A)A^2-\gamma_2(A)A]=\det(A)=a_1a_2a_3-(a_1b_1^2+a_2b_2^2+a_3b_3^2)\,.
}
Because $\Tr[A]=\Tr[A_1]$ but the determinant $\det(A)\le \det(A_1)$, the spectrum of $A_1$ is squashed compared to the spectrum of $A$.

Positive definiteness of $A$ is equivalent to the conditions
\eq{
\gamma_1(A)>0\,,\quad
\gamma_2(A)<0\,,\quad
\gamma_3(A)>0\,.\quad
}
If one eigenvalue of $A$ is zero, then $\gamma_3(A)=0$, and if two are zero, then $\gamma_2(A)=0$ as well.

For $A=Y^d{Y^d}^\dag$ positive definite, we can obtain an interesting formula for $\mu$ in
\eq{
\label{A2-tilde}
\La1^{-1/2}\La2\La1^{-1/2}\sim \mu\mtrx{0&&\cr &0&1\cr &-1&0}\,,
}
where the righthandside is the canonical form of a real antisymmetric matrix.
In the basis \eqref{basis:diag.A1}, the matrix in \eqref{A2-tilde} can be written as
\eq{
(\La1^{-1/2}\La2\La1^{-1/2})_{ij}=\eps_{ijk}\tb_k\,,
}
where
\eq{
\label{btilde.i}
\tb_i\equiv\frac{\sqrt{a_i}b_i}{\sqrt{a_1a_2a_3}}\,.
}
It is clear that $\mu=\sqrt{\tb_i\tb_i}$.
Rewriting the last relation in \eqref{gammai:A} as
\eq{
\det(A)=\det(A_1)(1-\tb_i\tb_i)\,,
}
we obtain the formula
\eq{
\label{mu.formula}
\mu=\sqrt{1-\frac{\det A}{\det A_1}}\,.
}

For $A=ww^\dag$ in \eqref{ww:xy}, we substitute $a_i\to x_i$ and $b_i\to -y_i$. If $A$ is rank 1, by orthogonal transformations, we can go to a basis where $A$ is block diagonal. Then $x_1=0$ and $y_2=y_3=0$. We are left with the down-right subblock nonzero. The zero determinant condition on this subblock leads to $y_1^2=x_2x_3$.
If $A$ is rank 2, the determinant $\det(A)$ in \eqref{gammai:A} is zero and
\eq{
\label{detaA=0}
x_1x_2x_3=x_1y_1^2+x_2y_2^2+x_3y_3^2\,.
}
We can still have $x_1=0$ as a special possibility, in which case, $y_2=y_3=0$.
But $\gamma_2(A)<0$ leads to $x_1^2<y_2y_3$.
If $A$ is rank 3, $\det(A)>0$ leads to 
\eq{
\label{detaA>0}
x_1x_2x_3>x_1y_1^2+x_2y_2^2+x_3y_3^2\,.
}
Equations \eqref{detaA=0} and \eqref{detaA>0} written in terms of $\ty_i$ in \eqref{def:ty1} leads to the ellipsoid condition \eqref{ellipsoid}.



\end{document}